\documentclass[prl,twocolumn,superscriptaddress,showpacs,aps,amsmath,amssymb]{revtex4-1}
\usepackage[toc]{appendix}
\usepackage[utf8]{inputenc}
\usepackage{url}
\usepackage{graphicx}
\usepackage{subfigure}
\usepackage{color}
\usepackage{xcolor}
\usepackage{epstopdf}
\usepackage{float}
\usepackage{ulem}

\usepackage{hyperref}
\usepackage{dcolumn}
\usepackage{bm}
\usepackage{psfrag}
\usepackage{array}
\usepackage{epsfig}
\usepackage{amsmath}
\usepackage{amssymb}
\usepackage{mathbbol}

\usepackage{titlesec}

\titleformat{\paragraph}
{\normalfont\normalsize\bfseries}{\theparagraph}{1em}{}
\titlespacing*{\paragraph}{0pt}{3.25ex plus 1ex minus .2ex}{1.5ex plus .2ex}

\newcommand{\beq}{\begin{equation}}
	\newcommand{\eeq}{\end{equation}}

\newcommand\ba{\begin{eqnarray}}
	\newcommand\ea{\end{eqnarray}}
\newcommand\be{\begin{equation}}
	\newcommand\ee{\end{equation}}

\def\dag{\dagger}

\begin{document}
	
	\title{Quantum critical behavior of entanglement in lattice bosons with cavity-mediated long-range interactions}
	\author{Shraddha Sharma \footnote{Electronic address: \texttt{shrdha1987@gmail.com}}}
	\thanks{Equal contribution}
	\affiliation{Theoretische Physik, Saarland University, Campus E2.6, 66123 Saarbr\"ucken, Germany}
	\affiliation{ICTP - The Abdus Salam International Center for Theoretical Physics, Strada Costiera 11, 34151 Trieste, Italy}
	\author{Simon B. J\"ager \footnote{Electronic address: \texttt{sjaeger@physik.uni-kl.de}}}
	\thanks{Equal contribution}
	\affiliation{Theoretische Physik, Saarland University, Campus E2.6, 66123 Saarbr\"ucken, Germany}
	 \affiliation{Physics Department and Research Center OPTIMAS, Technische Universit\"at Kaiserslautern, D-67663, Kaiserslautern, Germany}
	 	\affiliation{JILA and Department of Physics, University of Colorado, 440 UCB, Boulder, CO 80309, USA}
	\author{Rebecca Kraus}
	\affiliation{Theoretische Physik, Saarland University, Campus E2.6, 66123 Saarbr\"ucken, Germany}
	\author{Tommaso Roscilde}
	\affiliation{Univ Lyon, Ens de Lyon, CNRS, Laboratoire de Physique, F-69342 Lyon, France}
	\author{Giovanna Morigi}
	\affiliation{Theoretische Physik, Saarland University, Campus E2.6, 66123 Saarbr\"ucken, Germany}

	\begin{abstract}
	 We analyze the ground-state entanglement entropy of the extended Bose-Hubbard model with infinite-range interactions. This model describes the low-energy dynamics of ultracold bosons tightly bound to an optical lattice and dispersively coupled to a cavity mode. The competition between onsite repulsion and global { cavity-induced} interactions leads to a rich phase diagram, which exhibits superfluid, supersolid, and insulating (Mott and checkerboard) phases.  We use a slave-boson treatment of harmonic quantum fluctuations around the mean-field solution and calculate the entanglement entropy across the phase transitions. At commensurate filling, the insulator-superfluid transition is signalled by a singularity in the area-law scaling coefficient of the entanglement entropy,  that is similar to the one reported for the standard Bose-Hubbard model. Remarkably, at the  continuous $\mathbb{Z}_2$ superfluid-to-supersolid transition we find a {\it critical} logarithmic term, regardless of the filling. This behavior originates from the appearance of a roton mode in the excitation and entanglement spectrum, becoming gapless at the critical point, and it is characteristic of collective models. 
	 \end{abstract}
	
	\maketitle

{\it Introduction.} 
Entanglement measures play a special role in the low-temperature 
physics of quantum many-body systems, as they probe the existence and structure of quantum correlations \cite{Amico:2008}. Different entanglement measures have been discussed and applied to classify the emerging states of quantum matter~\cite{Amico:2008,Horodecki:2009}. Among them, the entanglement entropy (EE) captures the presence of bipartite entanglement in pure states: the scaling of the EE of a connected subsystem with its size exhibits universal properties ~\cite{Eisert:2010,Plenio:2005,Cramer:2007} probing {\it e.g.} the presence of conventional long-range order \cite{Metlitski:2011}, or of topological order \cite{Levin:2006,Kitaev:2006}. Singularities in the scaling behavior of the EE can mark in a universal way 
quantum phase transitions separating ordered from disordered phases \cite{Vidal:2007,Calabrese:2009,Frerot:2016}.  

		\begin{figure}[h!]
		\center
		\includegraphics[width=0.8\linewidth]{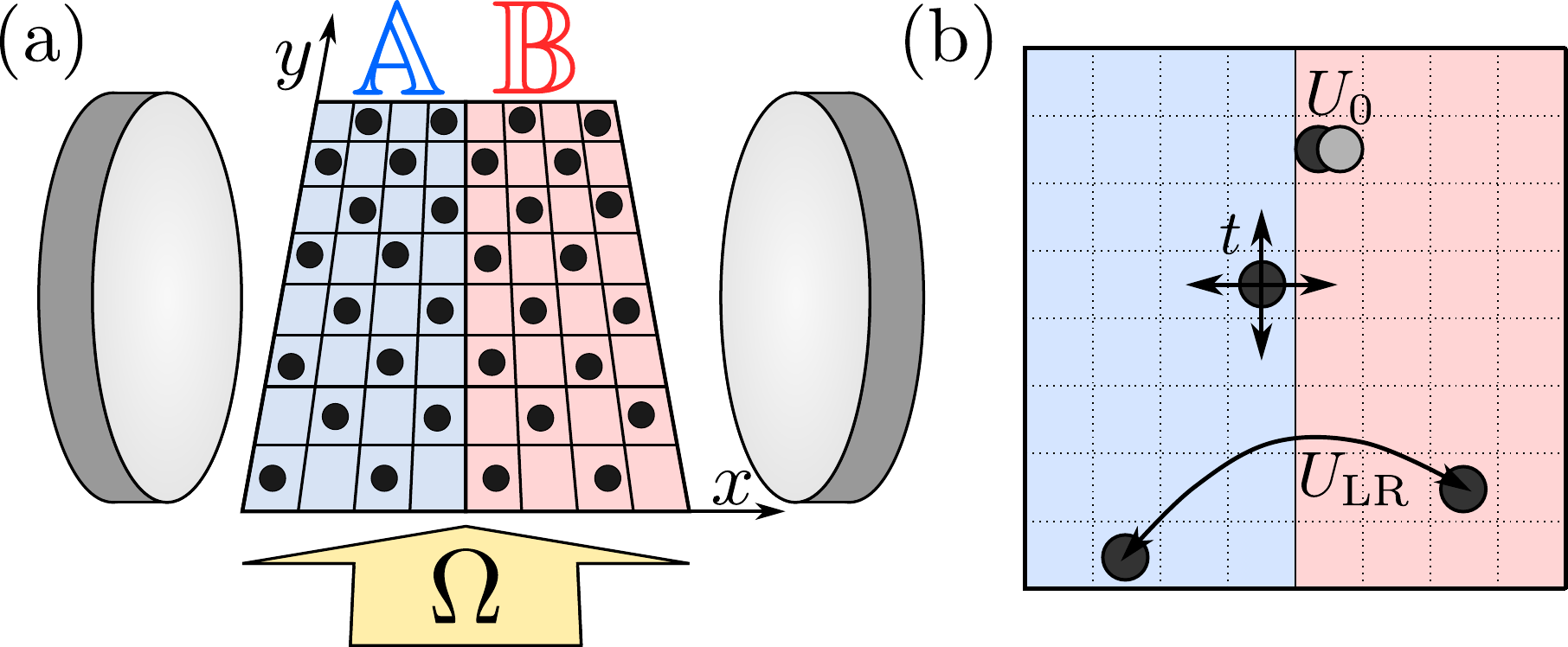}
		\caption{The Bose-Hubbard model with competing short and global interactions can be realized with atoms tightly bound by an optical lattice that coherently scatter laser photons ($\Omega$) into the mode of a high-finesse cavity~\cite{Landig:2016}. (a) The picture shows the geometry of the $\mathbb{A}$ / $\mathbb{B}$ 
		bipartition considered in this work. (b) Illustration of competing processes of the Hamiltonian: the nearest-neighbor tunneling (with amplitude $t$), the onsite repulsion ($U_0$), and the global density-density interactions ($U_\mathrm{LR}$), which are here attractive.} \label{Fig:1}
	\end{figure}

In this work we focus on the von Neumann EE, $S$, for a spatial bipartition $\mathbb{A}$ and $\mathbb{B}$ of an extended quantum system:
\begin{equation}
\label{Eq:EE}
S=-{\rm Tr}\{\rho_\mathbb{A}\log\rho_\mathbb{A}\}\,,
\end{equation}
where $\rho_\mathbb{A}={\rm Tr}_{\mathbb{B}}\{|\Psi_0\rangle\langle\Psi_0|\}$ is the density matrix obtained by tracing out the degrees of freedom of subsystem $\mathbb{B}$ from the ground state $|\Psi_0\rangle$.  Our purpose is to characterize the scaling of $S$ at continuous phase transitions resulting from competing short- and global-range interactions.

In fact, the interaction range can give rise to very different features. For short-range interactions the dominant scaling term of the EE is the so-called area-law term. This term grows with the size of the boundary between $\mathbb{A}$ and $\mathbb{B}$.  For a lattice of $d$ dimensions and $L$ lattice sites along each spatial dimensions the total number of lattice sites is $N=L^d$ and the EE scale as $L^{d-1}$ for a connected  subsystem $\mathbb{A}$ \cite{Eisert:2010}. The area-law scaling can be taken as an indication that quantum correlations between $\mathbb{A}$ and $\mathbb{B}$  involve primarily lattice sites close to the boundary \cite{Alba:2013}; yet,  for bosonic/spin systems in $d>1$ dimensions it persists even for ground states exhibiting critical or 
long-range correlations associated with the spontaneous breaking of a continuous symmetry \cite{Casini:2009,Hastingsetal2010,Humeniuk:2012}. In fact, criticality may lead at most to a singularity in the coefficient of the area-law scaling \cite{Singh:2012,Alba:2013, Helmes:2014,Frerot:2016},  while long-range correlations 
lead to the appearance of a universal subleading contribution to EE scaling. This contribution scales with the number of Goldstone modes $N_G$ as $(N_G/2)(d-1) \log L$ \cite{Metlitski:2011}. In contrast, for long-range interactions that decay with the inter-particle distance $r$ as $1/r^\alpha$ with $\alpha < d$, the geometric boundary between $\mathbb{A}$ and $\mathbb{B}$  becomes unimportant \cite{Campa:2009,Gong:2017,Kuwahara:2020}. For example, the area-law scaling disappears in the Dicke  \cite{Dicke:1973} and the Lipkin-Meshkov-Glick (LMG) model  \cite{Lipkin:1965,Meshkov:1965,Glick:1965}. Here, the ground state belongs to a symmetric subspace whose dimension grows linearly with $N$ and the EE scales as $\log(N)$ at the quantum critical point \cite{Vidal:2007}. To our knowledge, the scaling behavior of the EE is unexplored in the regime where short and global interactions compete.

{\it Extended Bose-Hubbard model.} In this Letter we analyze the scaling of the EE inthe two-dimensional extended Bose-Hubbard model of cavity quantum electrodynamics~\cite{Habibian:2013,Dogra:2016,Niederle:2016,Landig:2016}-- see Fig.~\ref{Fig:1} for a sketch. The Hamiltonian is the sum of the standard Bose-Hubbard Hamiltonian $\hat H_{\mathrm{BH}}$ \cite{Fisher:1989} and the cavity-mediated long-range interaction potential $\hat H_{\rm cav}$, namely, $\hat{H}=\hat H_{\mathrm{BH}}+\hat H_{\rm cav}$, with:
\begin{eqnarray}
&&\hat{H}_{\mathrm{BH}}=-t\sum_{\langle {\bm r},{\bm r'} \rangle}\hat{b}^{\dag}_{\bm r}\hat{b}_{{\bm r}'}+\sum_{\bm r}\left[\frac{U_0}{2}\hat{n}_{\bm r}(\hat{n}_{\bm r}-1)-\mu\hat{n}_{\bm r}\right]\,,\label{H:BH}\\
&&\hat H_{\rm cav}=-\frac{U_{\mathrm{LR}}}{N}\left[\sum_{{\bm r}} (-1)^{r_x+r_y}\hat{n}_{\bm r}\right]^2 \label{H:cav}\,,
\label{e.Ham}
\end{eqnarray}
where the parameters $t, U_0$, and $U_{\rm LR}$ are real and positive; $\hat{b}_{\bm r}^{\dag}$ ($\hat{b}_{\bm r}$)  create (annihilate) a boson at the site ${\bm r}=(r_x,r_y)$  of the square lattice; $\hat n_{\bm r}=\hat{b}_{\bm r}^\dagger \hat{b}_{\bm r}$ is the onsite density, and  $\langle {\bm r},{\bm r'} \rangle$ indicates a pair of nearest neighbors. In the following we assume periodic boundary conditions.  

\begin{figure}[t!]
			\center
			\includegraphics[width=1\linewidth]{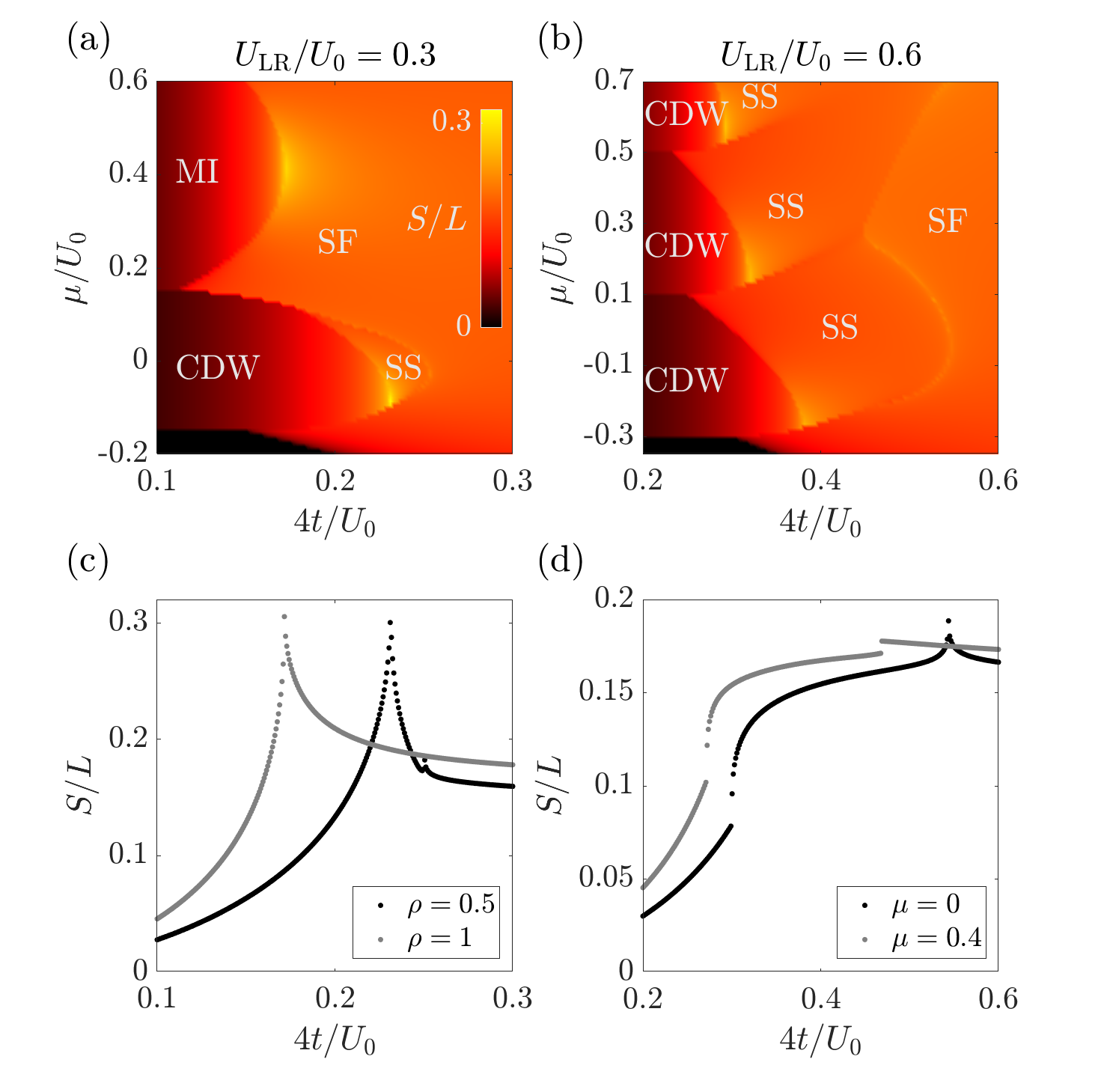}
			\caption {(color online) Color plot of the  half-system EE $S$ \eqref{Eq:EE} (see  the partition in Fig.~\ref{Fig:1}(a)) for (a) $U_\mathrm{LR}/U_0=0.3$ and  (b) $U_\mathrm{LR}/U_0=0.6$ as function of the tunneling $t$ and the chemical potential $\mu$ in units of $U_0$ for a $L\times L$ square lattice with $L=40$. The non-analyticities of $S$ coincide with the phase transition lines predicted by mean-field theory (not indicated here). The lower panels show $S/L$ as function of $t/U_0$  (c) at fixed density $\rho$  for $U_{\mathrm{LR}}=0.3U_0$; and (d) at fixed chemical potential $\mu/U_0$ for $U_{\mathrm{LR}}=0.6U_0$. Here the system the size is $L=60$.
				}\label{Fig:2}
		\end{figure}

Theoretical studies of the phase diagram of the Hamiltonian $\hat{H}$~\cite{Dogra:2016,Niederle:2016,Flottat:2017,Himbert:2019} reproduce the experimental results of Ref.~\cite{Landig:2016} for a cavity wavelength which is twice the periodicity of the optical lattice. This ground-state phase diagram features a rich palette of phases: The  nearest-neighbor hopping with amplitude $t$ favors the onset of superfluidity (SF)
while the onsite repulsion, with amplitude $U_0$, stabilizes a Mott insulator (MI) at commensurate filling. Global interactions, with amplitude $U_{\rm LR}$, induce a density modulation which supports scattering of photons into the cavity field. The  density modulation can result either in a charge density-wave (CDW) insulator, at integer or half-integer filling; or a supersolid (SS) phase, when it also exhibits superfluidity. Experimentally, the condensate fraction is revealed by time-of-flight measurements, while the onset of diagonal long-range order leads to the emission of coherent light at the cavity output~\cite{Landig:2016,Sierant:2019}.
 
The phase diagram is theoretically determined in the grand-canonical ensemble via the Gutzwiller mean-field (MF) approach \cite{Himbert:2019}. The ground state is written in the spatially factorized form 
$|\Psi_{0,{\rm MF}} \rangle= \otimes_{\bm r} |\psi_{{\bm r},0}\rangle$ with $ |\psi_{{\bm r},0}\rangle = \sum_{n=0}^{n_{\rm max}} f_{{\bm r},n}^{(0)} |n\rangle_{\bm r}$, where $|n\rangle_{\bm r}$ are the single-site Fock states and  $n_{\rm max}$ is a cutoff~\footnote{The cut-off $n_{\mathrm{max}}$ is chosen such that we find a negligible truncation error. For the results presented in this Letter this is achieved by using $n_{\rm max}=6$.}. The single-particle state $ |\psi_{{\bm r},0}\rangle$ is the ground state of the effective single-site Hamiltonian
$ \hat H^{\rm MF}_{\bm r}$, and it is determined self-consistently. Here, $ \hat H^{\rm MF}_{\bm r}=-zt\bar\varphi_{\bm r}(\hat{b}_{\bm r}+\hat{b}_{\bm r}^{\dagger}-\varphi_{\bm r})+\frac{U_0}{2}\hat{n}_{\bm r}(\hat{n}_{\bm r}-1) 
-\mu\hat{n}_{\bm r}-U_{\rm LR}\Theta (-1)^{r_x+r_y}\hat{n}_{\bm r}+NU_{\rm LR}\Theta^2/4$
where $z$ is the lattice coordination number ($z=4$) \cite{Himbert:2019}. Superfluidity 
is signalled by a non-vanishing value of the order parameter 
$\bar{\varphi}_{\bm r}=\sum_{{\bf r}'}\varphi_{{\bm r}'}/z$, where 
$\varphi_{\bm r}=\langle \hat{b}_{\bm r}\rangle$ and the sum runs over the nearest neighbors ${\bm r}'$ of ${\bm r}$. The onset of a density modulation is revealed by the order parameter $\Theta=2\langle\sum_{{\bm r}} (-1)^{r_x+r_y}\hat{n}_{\bm r}\rangle/N$. The upper panels of Fig.~\ref{Fig:2} display 
the phase diagram as a function of the ratios $t/U_0$ and $\mu/U_0$, the color scale gives the value of the EE, whose determination is discussed below. The subplots are evaluated for two values of the global potential $U_{\rm LR}$, chosen (a)~below and (b)~above the threshold $U_{\rm LR}^{\rm th}=U_0/2$, at which the MI phase becomes unstable. The non-analyticities of the EE coincide with the mean-field phase boundaries of Ref.~\cite{Himbert:2019}.   

The phase diagram  features first-order phase transitions and
three main types of {\it continuous} quantum phase transitions \cite{Dogra:2016,Flottat:2017,Himbert:2019}: { (type 1)} A commensurate $O(2)$ phase transition, separating either the MI from the SF at fixed integer density $\rho$, or the CDW from the SS at fixed integer density and at half filling. This transition occurs at the tip of the corresponding (MI or CDW) lobe; { (type 2)} A generic transition separating MI from SF and CDW from SS at incommensurate densities. This appears everywhere along the borders separating either MI/SF or CDW/SS, except for the lobe tips; { (type 3)} A continuous $\mathbb{Z}_2$ transition between the SS and the SF phase. { Some of these phase transitions change in fact from continuous to first-order} as the $t/U_0$ and $\mu/U_0$ ratios are tuned across the phase diagram. This is the case at the CDW/SS and SS/SF for small $t/U_0$, and at the SS/SF transition for  $\mu/U_0\gtrsim 0.25$ [see Fig.~\ref{Fig:2}(b)]. 
We note that  phase transitions changing from continuous to discontinuous as a function of the control fields (or the temperature) have also been reported for spin systems with competing short and long-range interactions~\cite{Campa:2009,Dauxois:2010}. 

{\it Slave-boson approach.}  We determine the entanglement properties in this rich phase diagram by including quantum correlations beyond the MF approximation. For this purpose we make use of a slave-boson approach \cite{Fresard:1994,Frerot:2016}, which we outline below and detail in the Supplemental Material (SM~\cite{SM}). Such an approach consists of using the full basis $\{|\psi_{{\bm r},\alpha}\rangle \}$ of eigenstates of $H^{\rm MF}_{\bm r}$ ($\alpha = 0, ..., n_{\rm max}$), and of defining associated slave-boson operators $\hat\gamma_{{\bm r},\alpha},~\hat\gamma^\dagger_{{\bm r},\alpha}$.  These operators fulfill the hardcore constraint $\sum_\alpha \hat\gamma_{\bm r, \alpha}^\dagger \hat\gamma_{\bm r, \alpha} = 1$ and are used to rewrite the original bosonic operators $\hat b_{\bm r} = \sum_{\alpha \beta} \sum_n  \sqrt{n} ~ f_{\bm r,n-1}^{(\alpha)} f_{\bm r,n}^{(\beta)} \hat\gamma^\dagger_{\bm r,\alpha} \hat\gamma_{\bm r,\beta}$ in $\hat{H}$. 
Within this formalism, the MF approximation corresponds to the condensation hypothesis of the ground-state slave bosons,  $\hat\gamma_{\bm r, 0}, \hat\gamma^\dagger _{\bm r, 0} = 1$ and $\langle   \hat\gamma_{\bm r, \alpha>0}^\dagger \hat\gamma_{\bm r, \alpha>0} \rangle=0$. The next level of approximation is to retain a finite population for the $\alpha>0$ bosons, and truncate the  full quartic Hamiltonian $\hat{H}$ to quadratic order in the $\hat\gamma_{\bm r, \alpha>0},  \hat\gamma^\dagger_{\bm r, \alpha>0}$ operators, by assuming that $\langle   \hat\gamma_{\bm r, \alpha>0}^\dagger \hat\gamma_{\bm r, \alpha>0}\rangle \ll 1$ and $\hat\gamma_{\bm r, 0}, \hat\gamma^\dagger _{\bm r, 0} \approx \left ( 1 -  \sum_{\alpha>0} \hat\gamma_{\bm r, \alpha}^\dagger \hat\gamma_{\bm r, \alpha}\right)^{1/2}$. The resulting Hamiltonian then reads $\hat{H} \approx \langle \Psi_{0,\rm MF} |  \hat{H} | \Psi_{0,\rm MF} \rangle + \hat{H}^{(2)}$ where  $\hat{H}^{(2)}$ is a quadratic form in the $\hat\gamma_{\bm r, \alpha>0}, \hat\gamma^\dagger_{\bm r, \alpha>0}$ operators~\cite{SM}. A Bogolyubov diagonalization of $\hat{H}^{(2)}$ reconstructs the quasi-particle spectrum $\omega_{\bm k, p}$ (where $p$ is a mode index, $p = 1, ..., n_{\rm max}$), and it allows us to calculate  the covariance matrix for subsystem $\mathbb{A}$, $\mathbf{C}_{\mathbb{A}}=\left[\mathbf{C}_{{\bm r},{\bm r}'}\right]_{{\bm r},{\bm r}'\in\mathbb{A}}$, where $\mathbf{C}_{{\bm r},{\bm r}'}=\langle \Psi_0|(\hat{\boldsymbol{\gamma}}_{\bm r},\hat{\boldsymbol{\gamma}}^{\dag}_{\bm r})^T(\hat{\boldsymbol{\gamma}}^{\dag}_{{\bm r}'},\hat{\boldsymbol{\gamma}}_{{\bm r}'})|\Psi_0\rangle$ with $\hat{\boldsymbol{\gamma}}_{\bm r}=(\hat\gamma_{{\bm r},1},\hat\gamma_{{\bm r},2},\ldots)$. For the remainder of this work, $\mathbb{A}$ will be the $L/2\times L$ rectangle obtained by cutting the $L\times L$ square lattice along the $y$ coordinate axis. The matrix $\mathbf{C}_{\mathbb{A}}$ contains all the information on the Gaussian reduced density matrix $\hat\rho_{\mathbb{A}}= e^{-\hat H_{\mathbb{A}}}$ for subsystem $\mathbb{A}$.  Operator $\hat{H}_{\mathbb{A}}$ is the so-called entanglement Hamiltonian, and it is a quadratic form in the $\hat\gamma_{\bm r, \alpha>0}, \hat\gamma^\dagger_{\bm r, \alpha>0}$. By means of a Bogolyubov transformation  $\hat{H}_{\mathbb{A}}$ becomes diagonal
\begin{align}
\hat{H}_{\mathbb{A}} = \sum_{k_y,m} \lambda_{k_y,m} \hat d^\dagger_{k_y,m} \hat d_{k_y,m}\,,
\end{align} where $\hat d_{k_y,m}, \hat d^\dagger_{k_y,m}$ are bosonic operators,  $\lambda_{k_y,m}$ represents the so-called (one-particle) entanglement spectrum, and we dropped a constant term. The  entanglement spectrum is labeled by the wavevector $k_y$ along the cut and by a further mode index $m$ associated with the motion perpendicular to the cut.  The EE $S$ corresponds then to the entropy of a gas of free bosons whose dispersion relation is the entanglement spectrum: $S = \sum_{k_y,m} s(n_{k_y,m})$ where $s(x) = (1+x)\log(1+x) - x \log x$ and $n_{k_y,m} = [\exp(\lambda_{k_y,m})-1]^{-1}$ is the Bose distribution. 

 {\it Entanglement phase diagram.}  Figures~\ref{Fig:2}(a) and (b) display $S$ in false colors throughout the phase diagrams. Remarkably, the EE exhibits characteristic signatures at {\it all} quantum phase transitions. In Fig.~\ref{Fig:2}(c) we report representative cuts at fixed density $\rho=1/2$ and $\rho=1$ for $U_{\rm LR}/U_0 = 0.3$. These cuts
show the existence of a sharp cusp singularity at the $O(2)$ MI/SF and CDW/SS transition (type 1). This singularity is associated with the appearance of a Higgs-like mode in the entanglement spectrum becoming gapless at the transition, and reflecting the softening of the Higgs mode in the quasi-particle spectrum \cite{Huber:2007}. The vanishing of the gap of the Higgs-like mode gives a singular contribution to the dominant, area-law scaling term. This behavior was reported in Ref.~\cite{Frerot:2016} for the MI/SF transition in the standard Bose-Hubbard model; and it also characterizes the CDW/SS transition, see SM~\cite{SM}. For the continuous generic MI/SF and CDW/SS transition (type 2), occurring away from the lobe tips in Fig.~\ref{Fig:2}(a) and (b), the EE singularity turns into a rounded maximum, similarly to the behavior of the standard Bose-Hubbard model~\cite{Frerot:2016}. In the extended Bose-Hubbard model, therefore, the critical behavior of entanglement at these phase transitions is due to the competition between hopping and contact short-range interactions.

 On the contrary, the long-range interactions play a crucial role for the continuous $\mathbb{Z}_2$ SS/SF transition (type 3) and its entanglement properties, as we argue below. We generally observe a cusp-like singularity of the EE at this transition. This is visible in the transition at fixed chemical potential ($\mu=0$) in Fig.~\ref{Fig:2}(d) as well as in the transition at constant density ($\rho = 1/2$) in Fig.~\ref{Fig:2}(c). In fact, the cusp singularity marks the {\it entire} SS/SF boundary whenever the corresponding transition is continuous. The robustness of this singularity in the EE makes the SS/SF transition stand out with respect to the MI/SF and the CDW/SS transitions of the same model and it represents our most important finding. Finally, when the transitions are first order, the EE exhibits a jump. This is visible in Fig.~\ref{Fig:2}(d) for the CDW/SS and the SS/SF transition at the constant value of the chemical potential $\mu=0.4U_0$.

\begin{figure}[t]
	\center
	\includegraphics[width=1\linewidth]{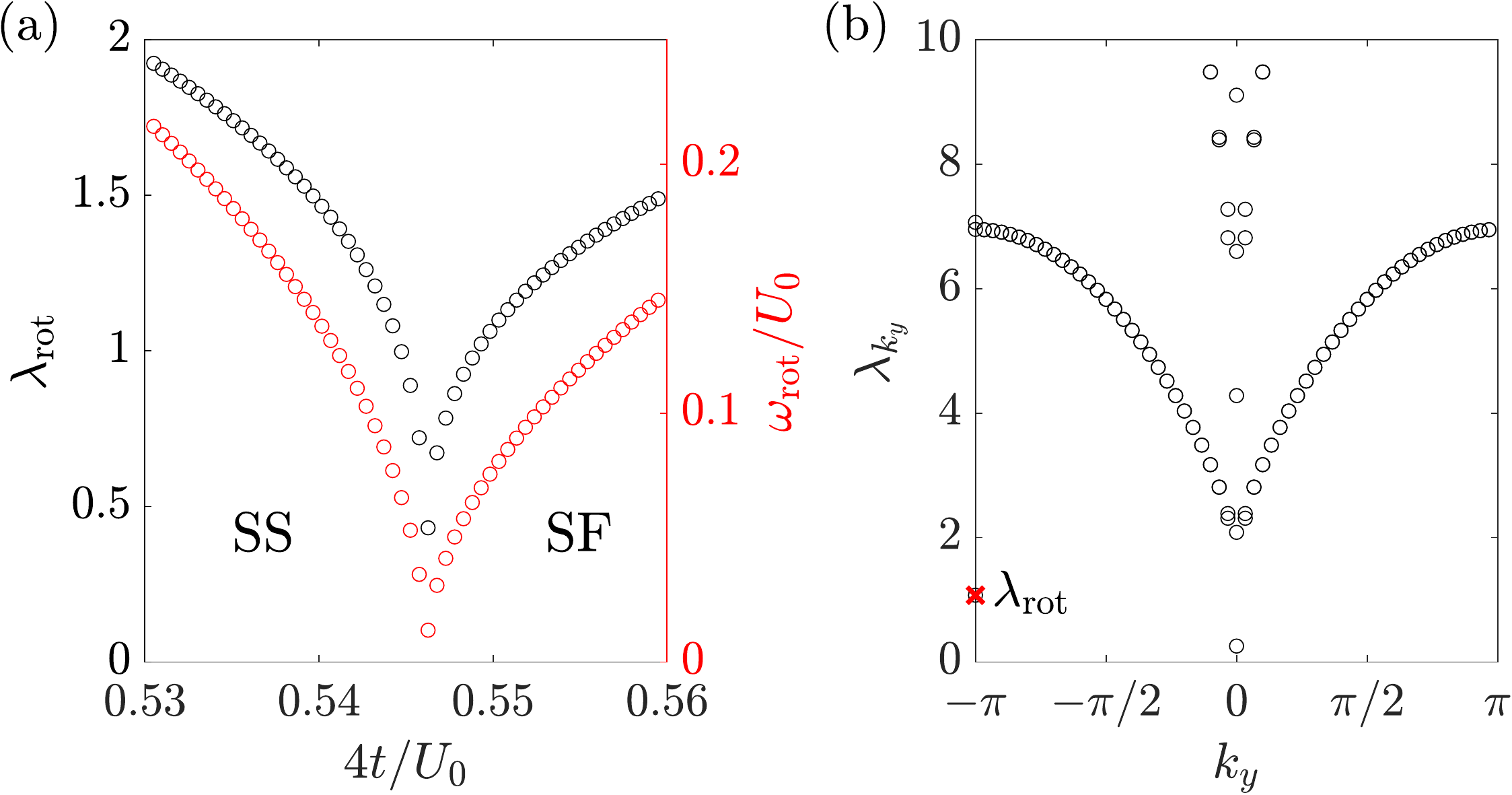}\
	\caption{(a) The roton-mode frequencies {$\lambda_{\rm rot} =  \lambda_{\pi,\bar m}$} ($\bar m$ being the index of roton-like mode) and $\omega_{\mathrm{rot}}$ in the entanglement and physical spectrum (respectively) as function of $4t/U_0$ across the SS-SF phase transition. Both frequencies vanish at the SS-SF transition. (b) Entanglement spectrum for $4t=0.55U_0$, $U_{\mathrm{LR}}=0.6U_0$, $\mu=-0.05U_0$ and $L=60$ as function of $k_y$. The roton mode in the  entanglement spectrum is highlighted by the red cross. }\label{Fig:3}\
\end{figure}

{\it Entanglement singularity from the roton mode.} To understand the origin of { the cusp singularity at the SS/SF transition, it is useful to} analyze the behavior of the excitation spectrum at the SS/SF transition.  The spectrum exhibits a vanishing gap { throughout} the SF and SS phase, namely the Goldstone mode related to the breaking of the $U(1)$ symmetry. Moreover, it is characterized by the critical softening of the roton frequency $\omega_{\rm rot}$ at wavevector ${\bm k_{\rm rot}} = (\pi,\pi)$ \cite{Dogra:2016}, which is the precursor of diagonal long-range order. The roton gap is displayed in Fig.~\ref{Fig:3}(a) as a function of $t/U_0$. { After closing at the SS/SF transition} it reopens in the SS phase: this is a consequence of elementary excitations of a $\mathbb{Z}_2$ crystal having a finite, non-vanishing gap just like in the CDW phase. The spectrum has a characteristic {\it dispersionless} and gapped structure in the vicinity of the critical roton mode at ${\bm k_{\rm rot}}$, reflecting { the fact} that the Fourier spectrum of the global interaction potential is a $\delta$-function at this wavevector. Correspondingly, in the entanglement spectrum [Fig.~\ref{Fig:3}(b)] a (boundary) roton-like mode, becomes gapless only at the SS/SF transition, and only for the frequency $\lambda_{\rm rot}$. This means that the EE acquires the critical roton contribution $S_{\rm rot} = s(n_{\rm rot}) \approx -\log \lambda_{\rm rot}$ as $\lambda_{\rm rot} \to 0$. 

\begin{figure}[t]
		\center
		\includegraphics[width=1\linewidth]{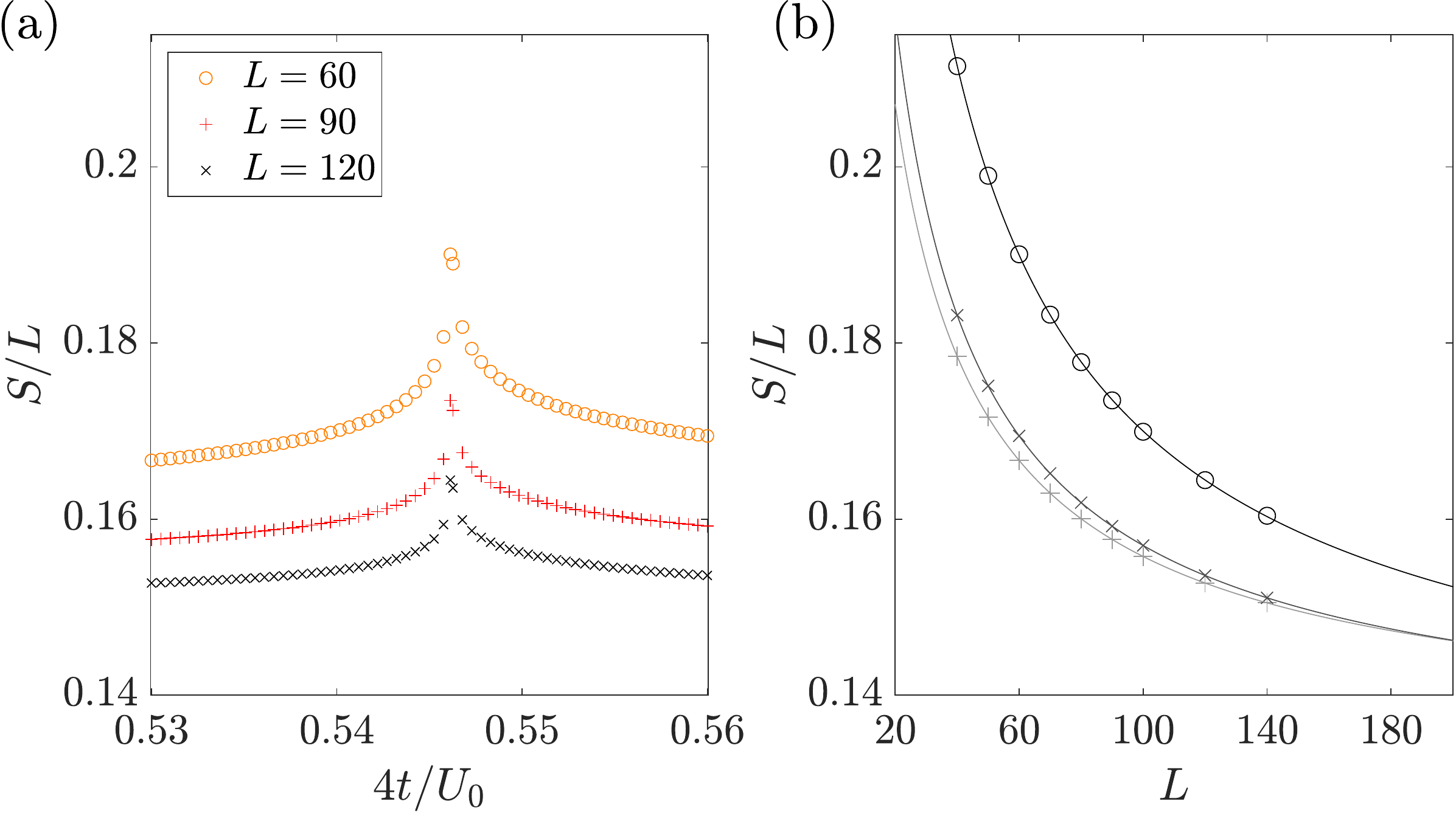}\
		\caption{ The half-system EE as function of the tunneling rate $t$ in units of $U_0$ across the transition from SS-SF, for $U_\mathrm{LR}=0.6U_0$ and constant $\mu=-0.05U_0$. (b) Scaling of the $S/L$ values at the maximum (``o" symbols); for $4t/U_0=0.56$ (``x" symbols); and for $4t/U_0=0.53$ (``+" symbols) for different $L$.   The coefficients $A$ and $B$ are obtained by fitting Eq.~\eqref{scal_ee} to $S$ vs $L$ and are given in the table.  For all the data in this figure the scaling exponent of the regularizing field $h(L)$ has been chosen as $\kappa = 4$.}\label{Fig:4}
	\end{figure}

 The scaling of the critical roton contribution with system size $L$ depends then on how the roton entanglement frequency $\lambda_{\rm rot}$ vanishes upon increasing $L$. This shall be handled with particular care. In fact, diagonalization of the quadratic Hamiltonian $\hat H^{(2)}$ leads to the unphysical result that the frequency $\omega_{\rm rot}$ vanishes for any finite system size at the SS/SF transition (and so does the frequency $\lambda_{\rm rot}$ of the entanglement spectrum). This is a common problem to the treatment of harmonic quantum fluctuations around a symmetry-breaking mean-field solution. In order to have meaningful finite-size results, we implement a regularization scheme  by applying a size-dependent field. This field couples to the order parameter and introduces a finite gap both in the excitation as well as in the entanglement spectrum \cite{Frerot:2015} \footnote{This treatment is required to get meaningful results for the EE in the SF phase. We apply a field term coupling directly to the bosonic field [see SM~\cite{SM}] and scaling like the $N^{-2}$ that introduces a gap in the excitation spectrum scaling as $N^{-1}$, in agreement with what is expected for systems breaking a $U(1)$ symmetry \cite{Frerot:2015}.}. For the $\mathbb{Z}_2$ critical point with infinite-range interactions, we add a term $- h(L)  \sum_{\bm r} (-1)^{\bm r} n_{\bm r}$ with $h(L) \sim L^{-\kappa}$. This choice is such that the gap introduced in the ``zero-modes'' mimics the scaling of the excitation gap at the transition $\omega_{\rm rot}\sim L^{-z}$, with $z$ the dynamical critical exponent.  The size-dependent field $h(L)$ also introduces a finite size scaling for the entanglement frequency $\lambda_{\rm rot}\sim L^{-\zeta}$. The determination of the scaling exponent $\kappa$ reproducing the correct $z$ exponent goes beyond the scope of our work. Yet, even though different power-law scalings of the applied field  lead to different scalings $\lambda_{\rm rot} \sim L^{-\zeta(\kappa)}$ for the roton mode,  all choices result in a singular logarithmic correction to the area law of the form $S_{\rm rot} \simeq \zeta \log L$.
  
{\it Scaling of the EE.}  
 We perform a scaling analysis of the half-system EE using the fitting function
\begin{align}
S= AL+B\log{L}+C\label{scal_ee}.
\end{align}  Figure~\ref{Fig:4} clearly shows that the spike in the EE at the transition is due to a spike in the fitted $B$ coefficient. This spike appears on top of the value $B \approx N_G(d-1)/2 = 1/2$ related to the contribution of the Goldstone mode, and is consistent with the singular logarithmic contribution to the EE coming from the roton mode. This is revealed by plotting $S/L$ vs. $L$: the curves at the critical point or away from it tend to a similar $A$ value -- the area-law scaling term -- but are offset sharply by the spike in the subleading term $ B \log L / L$. It is interesting to frame this result in the broader context of quantum phase transitions in models with global interactions \cite{Vidal:2007}. 
In the SM \cite{SM} we show that a slave-boson treatment 
of the LMG model
recovers exactly the $\log N$ scaling behavior of the EE at the critical point.  We relate this behavior quantitatively to the appearance of an isolated vanishing mode $\lambda_{\rm min}$ in an otherwise nearly dispersionless entanglement spectrum.

{\it Conclusions.}  We have shown that the entanglement entropy (EE) sheds light onto the role of the interaction range at the quantum phase transitions of the extended Bose-Hubbard model of cavity quantum electrodynamics (CQED).  The continuous phase transitions separating the insulating from the superfluid phases { exhibit a singularity in the coefficient of the area-law scaling of the EE}, as in the short-range Bose-Hubbard model. Remarkably, at the continuous $\mathbb{Z}_2$ superfluid/supersolid transition, the EE's behavior is 
is accompanied by a critical logarithmic scaling term of the EE, originating from the singular vanishing of the roton gap. The behavior of a vanishing gap in a dispersionless roton mode is 
 similar to the one reported at the quantum phase transition of collective spin models and is determined by the global-range potential. This analysis can be extended to characterize quantum phase transitions of driven-dissipative CQED models \cite{Jaeger:2019,Marino:2021,Deuar:2021}. The perspective of studying cavity-induced correlations in quantum gas microscopes \cite{Gross:2021} opens the possibility of measuring EE via the replica \cite{Islam:2015} or the random-measurement approach \cite{Brydges:2019} and it suggests that our predictions could be accessible to future experiments.

{\it Acknowledgements.} The authors are grateful to Ir\'en\'ee Fr\'erot, Lukas Himbert, and Astrid Elisabeth Niederle for discussions and helpful comments. This work has been supported by the Deutsche Forschungsgemeinschaft (DFG, German Research Foundation) via the priority program No. 1929 ``GiRyd'' and the CRC-TRR 306 ``QuCoLiMa'', Project-ID
No. 429529648, and by the German Ministry of Education and Research (BMBF) via the QuantERA project NAQUAS. Project NAQUAS has received funding from the QuantERA ERA-NET Cofund in Quantum Technologies implemented within the European Union's Horizon 2020 program. SBJ acknowledges support from the NSF Q-SEnSE Grant No. OMA 2016244; NSF PFC Grant No. 1734006. TR acknowledges support from ANR (EELS project) and QuantERA (MAQS project).  
\bibliography{arXivEELRBHM.bib}
\newpage
\onecolumngrid
\newpage
\setcounter{equation}{0}
\setcounter{figure}{0}

\renewcommand*{\citenumfont}[1]{S#1}
\renewcommand*{\bibnumfmt}[1]{[S#1]}
\renewcommand{\thesection}{S~\arabic{section}}
\renewcommand{\thesubsection}{\thesection.\arabic{subsection}}
\makeatletter 
\def\tagform@#1{\maketag@@@{(S\ignorespaces#1\unskip\@@italiccorr)}}
\makeatother
\makeatletter
\makeatletter \renewcommand{\fnum@figure}
{\figurename~S\thefigure}
\begin{center}
	\textbf{\large Supplemental Material: Quantum critical behavior of entanglement in lattice bosons with cavity-mediated long-range interactions}
	\end{center}
\twocolumngrid

	\tableofcontents
	\section{Derivation of the quadratic Hamiltonian}
	In this section we explain how we derive the quadratic Hamiltonian that describes fluctuations around the mean-field solution.
	
	\subsection{Quadratic Hamiltonian in real space}
	The local mean-field theory allows us to obtain  the mean-field ground state $|\psi_{{\bm r},0}\rangle$ at every site ${\bm r}$ of a square lattice
	\begin{align}
	\mathbb{L}=\{m{\bm e}_x+n{\bm e}_y\,|\,m,n=-L/2,-L/2+1,\dots,L/2-1\},\nonumber
	\end{align}
	with ${\bm e}_x=(1,0)^T$, ${\bm e}_y=(0,1)^T$, 
	as well as  the excited states $|\psi_{{\bm r},\alpha\geq1}\rangle$. In order to add correlations in the local mean-field Hamiltonian,  we use the slave-boson annihiliation and creation operators $\hat{\gamma}_{{\bm r},\alpha}$ and $\hat{\gamma}^{\dag}_{{\bm r},\alpha}$, respectively,  that annihilate or create the state $|\psi_{{\bm r},\alpha}\rangle$ at site ${\bm r}$. Those creation and annihilation operators obey the canonical commutation relations $[\hat{\gamma}_{{\bm r},\alpha},\hat{\gamma}_{{\bm r}',\beta}^{\dag}]=\delta_{{\bm r},{\bm r}'}\delta_{\alpha, \beta}$ and $[\hat{\gamma}_{{\bm r},\alpha},\hat{\gamma}_{{\bm r}',\beta}]=0$. They must satisfy a hard-core constraint $\sum_{\alpha}\hat{\gamma}^{\dag}_{{\bm r},\alpha}\hat{\gamma}_{{\bm r},\alpha}=1$ for the state at site $\bm r$ to be well defined. 
	
	Using this definition we can write down any local operator $\hat{O}_{\bm r}$ as
	\begin{align}
	\hat{O}_{\bm r}\rightarrow&\sum_{\alpha,\beta}\hat{\gamma}^{\dag}_{\bm r,\alpha}O_{{\bm r},\alpha\beta}\hat{\gamma}_{\bm r,\beta}\nonumber\\  =& O_{{\bm r},00}+\sum_{\alpha>0}\left(\hat{\gamma}^{\dag}_{{\bm r},0}O_{{\bm r},0\alpha}\hat{\gamma}_{\bm r,\alpha}+\hat{\gamma}^{\dag}_{{\bm r},\alpha}O_{{\bm r},\alpha0}\hat{\gamma}_{{\bm r},0}\right)\nonumber\\
	&+ \sum_{\alpha,\beta>0}\hat{\gamma}^{\dag}_{\bm r,\alpha}\left[O_{{\bm r},\alpha\beta}-O_{{\bm r},00}\delta_{\alpha,\beta}\right]\hat{\gamma}_{\bm r,\beta},\label{Eq:Or}
	\end{align}
	with $O_{{\bm r},\alpha\beta}=\langle\psi_{\bm r,\alpha}|\hat{O}_{\bm r}|\psi_{\bm r,\beta}\rangle$.  Here we used the fact that $\hat{\gamma}^{\dag}_{{\bm r},0}\hat{\gamma}_{{\bm r},0}=1-\hat{\epsilon}_{\bm r}$ with $\hat{\epsilon}_{{\bm r}}=\sum_{\alpha>0}\hat{\gamma}^{\dag}_{{\bm r},\alpha}\hat{\gamma}_{{\bm r},\alpha}$. In what follows we will now use the relation of Eq.~\eqref{Eq:Or} in every term of the Hamiltonian $\hat{H}$ of the main text and  keep fluctuations terms up to order $\hat{\epsilon}_{{\bm r}}$ but discard  terms which are beyond linear in $\hat\epsilon$. 
	
	As a result we obtain the Hamiltonian in the form
	\begin{align}
	\hat{H}=\hat{H}^{(0)}+\hat{H}^{(1)}+\hat{H}^{(2)},\label{HSB}
	\end{align}
	where $\hat{H}^{(i)}$ collects all terms of the order of $\hat{\epsilon}_{{\bm r}}^{i/2}$ with $i=0,1,2$. We will now derive the individual terms in detail.
	
	The zeroth order term $\hat{H}^{(0)}$ in Eq.~\eqref{HSB} is obtained by simply substituting all local operators  with their local mean-field value as visible in the first term of Eq.~\eqref{Eq:Or}. Therefore this term is given by the mean-field energy, and it reads  
	\begin{align}
	\hat{H}^{(0)}= -t\sum_{<{\bm r},{\bm r'}>}\varphi^{*}_{\bm r}\varphi_{{\bm r}'}+\sum_{{\bm r}\in\mathbb{L}}H_{{\bm r},00}-U_{\mathrm{LR}}N\Theta^2/4, \label{H(0)} 
	\end{align}
	with the on-site Hamiltonian
	\begin{align}
	\hat{H}_{\bm r}=\frac{U_0}{2}\hat{n}_{\bm r}(\hat{n}_{\bm r}-1)-\mu\hat{n}_{\bm r},\label{Hloc}
	\end{align}
	and $\varphi_{\bm r}=b_{{\bm r},00}$.
	The first-order term $\hat{H}^{(1)}$ in Eq.~\eqref{HSB} corresponds to using the $\sum_{\alpha>0}$ term in Eq.~\eqref{Eq:Or} for one of the operators appearing in the Hamiltonian. Using the same notation as before we can write the first-order term of the Bose-Hubbard Hamiltonian (namely, without the cavity-induced interactions) as
	\begin{align*}
	\hat{H}_{\mathrm{BH}}^{(1)}=&-4t\sum_{{\bm r}\in\mathbb{L}}\bar\varphi_{\bm r}^*\sum_{\alpha>0}\left(\hat{\gamma}^{\dag}_{{\bm r},0}b_{{\bm r},0\alpha}\hat{\gamma}_{\bm r,\alpha}+\hat{\gamma}^{\dag}_{{\bm r},\alpha}b_{{\bm r},\alpha0}\hat{\gamma}_{{\bm r},0}\right)\\
	&-4t\sum_{{\bm r}\in\mathbb{L}}\bar\varphi_{\bm r}\sum_{\alpha>0}\left(\hat{\gamma}^{\dag}_{{\bm r},0}b^{\dag}_{{\bm r},0\alpha}\hat{\gamma}_{\bm r,\alpha}+\hat{\gamma}^{\dag}_{{\bm r},\alpha}b^{\dag}_{{\bm r},\alpha0}\hat{\gamma}_{{\bm r},0}\right)\\
	&+\sum_{{\bm r}\in\mathbb{L}}\sum_{\alpha>0}\left(\hat{\gamma}^{\dag}_{{\bm r},0}H_{{\bm r},0\alpha}\hat{\gamma}_{\bm r,\alpha}+\hat{\gamma}^{\dag}_{{\bm r},\alpha}H_{{\bm r},\alpha0}\hat{\gamma}_{{\bm r},0}\right).
	\end{align*}
	Here, we used the definition $\bar{\varphi}_{\bm r}=\sum_{{\bm r'}\in\mathcal{N}({\bm r})}\varphi_{{\bm r}'}/4$, where $\mathcal{N}({\bm r})=\{{\bm r}\pm{\bm e}_x,{\bm r}\pm{\bm e}_y\}$. Notice that with this definition we can switch  between $\sum_{<{\bm r},{\bm r}'>}\leftrightarrow\sum_{{\bm r}\in\mathbb{L}}\sum_{{\bm r}'\in\mathcal{N}({\bm r})}$.
	
	Equivalently we can define $\hat{H}_{\mathrm{cav}}^{(1)}$ (containing the cavity-induced interactions) that takes the form
	\begin{align*}
	\hat{H}_{\mathrm{cav}}^{(1)}=&-U_\mathrm{LR}\Theta\sum_{{\bm r}\in\mathbb{L}}\sum_{\alpha>0}Z_{\bm r}\left(\hat{\gamma}^{\dag}_{{\bm r},0}n_{{\bm r},0\alpha}\hat{\gamma}_{\bm r,\alpha}+\hat{\gamma}^{\dag}_{{\bm r},\alpha}n_{{\bm r},\alpha0}\hat{\gamma}_{{\bm r},0}\right)
	\end{align*}
	with $Z_{\bm r}=(-1)^{r_x+r_y}$.
	Combining the two equations for $\hat{H}_{\mathrm{BH}}^{(1)}$ and $\hat{H}_{\mathrm{cav}}^{(1)}$ leads to the result 
	\begin{align}
	\hat{H}^{(1)}=\sum_{{\bm r}\in\mathbb{L}}\sum_{\alpha>0}\left(\hat{\gamma}^{\dag}_{{\bm r},0}H_{{\bm r},0\alpha}^{\rm MF}\hat{\gamma}_{\bm r,\alpha}+\mathrm{H.c.}\right) \label{H(1)},  
	\end{align}
	where we defined the mean-field Hamiltonian
	\begin{eqnarray}\label{hmf2}
	H^{\rm MF}_{{\bm r}}&=&-4t(\bar\varphi_{\bm r}^*\hat{b}_{\bm r}+\hat{b}_{\bm r}^{\dagger}\bar\varphi_{\bm r})+\hat{H}_{\bm r}-U_\mathrm{LR} \Theta  Z_{\bm r}  \hat{n}_{\bm r}.\nonumber \label{hmf}\\
	\end{eqnarray}
	This first order term vanishes due to the choice of our eigenbasis $|\psi_{{\bm r},\alpha}\rangle$, made of eigenstates of the mean-field Hamiltonian in Eq.~\eqref{hmf}.
	Consequently we find
	\begin{align}
	H_{{\bm r},\alpha\beta}^{\rm MF}=\langle\psi_{\bm r,\alpha}|\hat{H}_{{\bm r}}^{\rm MF}|\psi_{\bm r,\beta}\rangle=E_{{\bm r},\alpha}^{\mathrm{MF}}\delta_{\alpha\beta}, \label{Emeanfield}   
	\end{align}
	with mean-field eigenenergies $E_{{\bm r},\alpha}^{\mathrm{MF}}$. Because of the orthogonality of the eigenstates we immediately obtain $\hat{H}^{(1)}=0$.
	
	To derive the second order term $\hat H^{(2)}$ in Eq.~\eqref{HSB}, we must consider  (i) products of second-order and zeroth-order terms (as in Eq.~\eqref{Eq:Or})  and  (ii) products of two first-order terms. We will refer to those combinations by $\hat{H}_{\mathrm{BH}}^{(2),(i)}$,  $\hat{H}_{\mathrm{BH}}^{(2),(ii)}$, and $\hat{H}_{\mathrm{cav}}^{(2),(i)}$,  $\hat{H}_{\mathrm{cav}}^{(2),(ii)}$, respectively. We start by calculating
	\begin{align*}
	\hat{H}_{\mathrm{BH}}^{(2),(i)}=&-4t\sum_{{\bm r}\in\mathbb{L}}\bar\varphi_{\bm r}^*\sum_{\alpha,\beta>0}\hat{\gamma}^{\dag}_{\bm r,\alpha}\left[b_{{\bm r},\alpha\beta}-b_{{\bm r},00}\delta_{\alpha,\beta}\right]\hat{\gamma}_{\bm r,\beta}\\
	&-4t\sum_{{\bm r}\in\mathbb{L}}\bar\varphi_{\bm r}\sum_{\alpha,\beta>0}\hat{\gamma}^{\dag}_{\bm r,\alpha}\left[b^{\dag}_{{\bm r},\alpha\beta}-b^{\dag}_{{\bm r},00}\delta_{\alpha,\beta}\right]\hat{\gamma}_{\bm r,\beta}\\
	&+\sum_{{\bm r}\in\mathbb{L}}\sum_{\alpha,\beta>0}\hat{\gamma}^{\dag}_{\bm r,\alpha}\left[H_{{\bm r},\alpha\beta}-H_{{\bm r},00}\delta_{\alpha,\beta}\right]\hat{\gamma}_{\bm r,\beta}.
	\end{align*}
	and
	\begin{align*}
	\hat{H}_{\mathrm{cav}}^{(2),(i)}=&-U_\mathrm{LR}\Theta \sum_{{\bm r}\in\mathbb{L}}\sum_{\alpha,\beta>0}Z_{\bm r}\hat{\gamma}^{\dag}_{\bm r,\alpha}\left[n_{{\bm r},\alpha\beta}-n_{{\bm r},00}\delta_{\alpha,\beta}\right]\hat{\gamma}_{\bm r,\beta}.
	\end{align*}
	Combining those two terms results in
	\begin{align*}
	\hat{H}_{\mathrm{BH}}^{(2),(i)}+ \hat{H}_{\mathrm{cav}}^{(2),(i)}=\sum_{{\bm r}\in\mathbb{L}}  \sum_{\alpha,\beta>0}\hat{\gamma}_{{\bm r},\alpha}^{\dag} A^{(0)}_{{\bm r}\alpha,{\bm r}\beta}\hat{\gamma}_{{\bm r},\beta},
	\end{align*}
	where we used Eq.~\eqref{Emeanfield} and defined 
	\begin{eqnarray}
	A^{(0)}_{{\bm r}\alpha,{\bm r}\beta}&=&\delta_{\alpha,\beta}(E_{{\bm r},\alpha}^{\textrm{MF}}-E_{{\bm r},0}^{\textrm{MF}}).\label{A01}
	\end{eqnarray}
	
	In a next step we calculate 
	\begin{align*}
	\hat{H}_{\mathrm{BH}}^{(2),(ii)}=&-t\sum_{<{\bm r},{\bm r'}>}\sum_{\alpha,\beta>0}\hat{\gamma}^{\dag}_{{\bm r},\alpha}b^{\dag}_{{\bm r},\alpha0}\hat{\gamma}_{{\bm r},0}\hat{\gamma}^{\dag}_{{\bm r}',0}b_{{\bm r}',0\beta}\hat{\gamma}_{{\bm r}',\beta}\\
	&-t\sum_{<{\bm r},{\bm r'}>}\sum_{\alpha,\beta>0}\hat{\gamma}^{\dag}_{{\bm r},0}b^{\dag}_{{\bm r},0\alpha}\hat{\gamma}_{{\bm r},\alpha}\hat{\gamma}^{\dag}_{{\bm r}',0}b_{{\bm r}',0\beta}\hat{\gamma}_{{\bm r}',\beta}\\
	&-t\sum_{<{\bm r},{\bm r'}>}\sum_{\alpha,\beta>0}\hat{\gamma}^{\dag}_{{\bm r},\alpha}b^{\dag}_{{\bm r},\alpha0}\hat{\gamma}_{{\bm r},0}\hat{\gamma}^{\dag}_{{\bm r}',\beta}b_{{\bm r}',\beta0}\hat{\gamma}_{{\bm r}',0}\\
	&-t\sum_{<{\bm r},{\bm r'}>}\sum_{\alpha,\beta>0}\hat{\gamma}^{\dag}_{{\bm r},0}b^{\dag}_{{\bm r},0\alpha}\hat{\gamma}_{{\bm r},\alpha}\hat{\gamma}^{\dag}_{{\bm r}',\beta}b_{{\bm r}',\beta0}\hat{\gamma}_{{\bm r}',0}.
	\end{align*}
	Now using $\hat{\gamma}_{{\bm r},0}\approx1$, since corrections would be at least of third order in $\hat \gamma_{\alpha>0}$, we can write
	\begin{align*}
	\hat{H}_{\mathrm{BH}}^{(2),(ii)}=\sum_{{\bm r},{\bm r'}\in\mathbb{L}}
	\sum_{\alpha,\beta>0}\bigg[& \hat{\gamma}_{{\bm r},\alpha}^{\dag} A^{\mathrm{hop}}_{{\bm r}\alpha,{\bm r}'\beta}\hat{\gamma}_{{\bm r}',\beta}\nonumber\\
	&+\frac{1}{2}(\gamma_{{\bm r}\alpha}B^{\mathrm{hop}}_{{\bm r}\alpha,{\bm r}'\beta}\gamma_{{\bm r}',\beta}+\mathrm{H.c.})\bigg],
	\end{align*}
	where we have the symmetrized form
	$$b^{\dag}_{{\bm r},0\alpha}b_{{\bm r}',0\beta} \leftrightarrow \frac{b^{\dag}_{{\bm r},0\alpha}b_{{\bm r}',0\beta}+b^{\dag}_{{\bm r}',0\beta}b_{{\bm r},0\alpha}}{2}$$
	since $\hat{\gamma}_{\bm r,\alpha}$ and $\hat{\gamma}_{{\bm r}',\beta}$ commute, and the definitions 
	\begin{eqnarray}
	A^{\mathrm{hop}}_{{\bm r}\alpha,{\bm r}'\beta}&=&-t(b_{{\bm r},\alpha0}^\dagger b_{{\bm r}',0\beta}+{b_{{\bm r},\alpha0}b_{{\bm r}',0\beta}^\dagger})\delta_{{\bm r}',\mathcal{N}({\bm r})},\label{A11}\\
	B^{\mathrm{hop}}_{{\bm r}\alpha,{\bm r}'\beta}&=&-t(b_{{\bm r},0\alpha}^\dagger b_{{\bm r}',0\beta}+{b_{{\bm r},0\alpha}b_{{\bm r}',0\beta}^\dagger})\delta_{{\bm r}',\mathcal{N}({\bm r})}.\label{Bhop}
	\end{eqnarray}
	Here, we have used the notation $\delta_{{\bm r}',\mathcal{N}({\bm r})}=1$ if ${\bm r}'\in\mathcal{N}({\bm r})$ (nearest-neighbour) and $\delta_{{\bm r}',\mathcal{N}({\bm r})}=0$ otherwise.
	
	For the last term we calculate
	\begin{align*}
	\hat{H}_{\mathrm{cav}}^{(2),(ii)}=&-\frac{U_\mathrm{LR}}{N}\sum_{{\bm r},{\bm r'}\in\mathbb{L}}\sum_{\alpha,\beta>0}Z_{\bm r}Z_{{\bm r}'}\hat{\gamma}^{\dag}_{{\bm r},\alpha}n_{{\bm r},\alpha0}n_{{\bm r}',0\beta}\hat{\gamma}_{{\bm r}',\beta}\\
	&-\frac{U_\mathrm{LR}}{N}\sum_{{\bm r},{\bm r'}\in\mathbb{L}}\sum_{\alpha,\beta>0}Z_{\bm r}Z_{{\bm r}'}n_{{\bm r},0\alpha}\hat{\gamma}_{{\bm r},\alpha}n_{{\bm r}',0\beta}\hat{\gamma}_{{\bm r}',\beta}\\
	&-\frac{U_\mathrm{LR}}{N}\sum_{{\bm r},{\bm r'}\in\mathbb{L}}\sum_{\alpha,\beta>0}Z_{\bm r}Z_{{\bm r}'}\hat{\gamma}^{\dag}_{{\bm r},\alpha}n_{{\bm r},\alpha0}\hat{\gamma}^{\dag}_{{\bm r}',\beta}n_{{\bm r}',\beta0}\\
	&-\frac{U_\mathrm{LR}}{N}\sum_{{\bm r},{\bm r'}\in\mathbb{L}}\sum_{\alpha,\beta>0}Z_{\bm r}Z_{{\bm r}'}n_{{\bm r},0\alpha}\hat{\gamma}_{{\bm r},\alpha}\hat{\gamma}^{\dag}_{{\bm r}',\beta}n_{{\bm r}',\beta0},
	\end{align*}
	where we have already used $\hat{\gamma}_{{\bm r},0}\approx1$. This last term can be written as
	\begin{align*}
	\hat{H}_{\mathrm{cav}}^{(2),(ii)}=\sum_{{\bm r},{\bm r'}\in\mathbb{L}}\sum_{\alpha,\beta>0}\bigg[& \hat{\gamma}_{{\bm r},\alpha}^{\dag} A^{\mathrm{LR}}_{{\bm r}\alpha,{\bm r}'\beta}\hat{\gamma}_{{\bm r}',\beta}\nonumber\\
	&+\frac{1}{2}(\gamma_{{\bm r}\alpha}B^{\mathrm{LR}}_{{\bm r}\alpha,{\bm r}'\beta}\gamma_{{\bm r}',\beta}+\mathrm{H.c.})\bigg],
	\end{align*}
	with \begin{eqnarray}
	A^{\mathrm{LR}}_{{\bm r}\alpha,{\bm r}'\beta}&=&-\frac{2U_\mathrm{LR}}{N}Z_{\bm r}Z_{{\bm r}'}n_{{\bm r},\alpha0}n_{{\bm r}',0\beta},\label{ALR}\\
	B^{\mathrm{LR}}_{{\bm r}\alpha,{\bm r}'\beta}&=&-\frac{2U_\mathrm{LR}}{N}Z_{\bm r}Z_{{\bm r}'}n_{{\bm r},0\alpha}n_{{\bm r}',0\beta}.\label{BLR}
	\end{eqnarray}
	Adding all terms, $\hat{H}_{\mathrm{BH}}^{(2),(i)}$,$\hat{H}_{\mathrm{cav}}^{(2),(i)}$, $\hat{H}_{\mathrm{BH}}^{(2),(ii)}$, and $\hat{H}_{\mathrm{cav}}^{(2),(ii)}$, leads to the result 
	\begin{align}
	\hat{H}^{(2)}=\sum_{{\bm r},{\bm r'}\in\mathbb{L}}\sum_{\alpha,\beta>0}\bigg[& \hat{\gamma}_{{\bm r},\alpha}^{\dag} A_{{\bm r}\alpha,{\bm r}'\beta}\hat{\gamma}_{{\bm r}',\beta}\nonumber\\
	&+\frac{1}{2}(\gamma_{{\bm r}\alpha}B_{{\bm r}\alpha,{\bm r}'\beta}\gamma_{{\bm r}',\beta}+\mathrm{H.c.})\bigg]
	\label{H2}	
	\end{align}
	
	\noindent with
	
	\begin{eqnarray}\label{Aab1}
	A_{{\bm r}\alpha,{\bm r}'\beta}&=&A^{(0)}_{{\bm r}\alpha,{\bm r}'\beta}\delta_{{\bm r},{\bm r}'}+A^{(1)}_{{\bm r}\alpha,{\bm r}'\beta},\\
	A^{(1)}_{{\bm r}\alpha,{\bm r}'\beta}&=&A^{\mathrm{hop}}_{{\bm r}\alpha,{\bm r}'\beta}+A^{\mathrm{LR}}_{{\bm r}\alpha,{\bm r}'\beta},\\
	B_{{\bm r}\alpha,{\bm r}'\beta}&=&B^{\mathrm{hop}}_{{\bm r}\alpha,{\bm r}'\beta}+B^{\mathrm{LR}}_{{\bm r}\alpha,{\bm r}'\beta}.
	\end{eqnarray}
	\noindent
	We emphasize that, although this general notation suggests that all matrices denoted by $A$ and $B$ can be different for different ${\bm r,\bm r}'$, they depend in our case only on whether ${\bm r, \bm r}'$ are ``even'' and/or ``odd''. Therefore for a local operator $\hat{O}_{\bm r}$ we can define
	\begin{align}
	O_{{\bm r},\alpha\beta}  =& O_{\mathrm{e},\alpha\beta}, \text{ if }{\bm r}\in\mathbb{L}^+\nonumber\\
	O_{{\bm r},\alpha\beta} =& O_{\mathrm{o},\alpha\beta}, \text{ if }{\bm r}\in\mathbb{L}^-, \label{evenodd}
	\end{align}
	where
	\begin{align}
	\mathbb{L}^\pm=\{{\bm r}\in\mathbb{L}\,|\,Z_{\bm r}=\pm1\}.
	\end{align}

	\subsection{Quadratic Hamiltonian in $\boldmath k$-space}
	In this section we will derive the form of the quadratic Hamiltonian in ${\bm k}$-space.  To this scope we define the Brillouin zone $\mathbb{B}=\{\frac{2\pi}{L}m{\bm e}_x+\frac{2\pi}{L}n{\bm e}_y\,|\,m,n=-L/2,-L/2+1\dots,L/2-1\}$ and the operators
	\begin{eqnarray}\label{g1}
	\gamma_{{\bm k},\alpha}&=&\frac{1}{\sqrt{N}}\sum_{{\bm r}\in\mathbb{L}} e^{-i{\bm k}\cdot{\bm r}}\gamma_{{\bm r},\alpha},\nonumber\\
	\gamma_{{\bm r},\alpha}&=&\frac{1}{\sqrt{N}}\sum_{{\bm k}\in\mathbb{B}} e^{i{\bm k}\cdot{\bm r}}\gamma_{{\bm k},\alpha}.
	\end{eqnarray}
	
	In the following we will use the fact that the matrix elements $A_{{\bm r}\alpha,{\bm r}'\beta}$ and $B_{{\bm r}\alpha,{\bm r}'\beta}$ do not depend explicitly on the pair ${\bm r},{\bm r}'$ but solely on whether they correspond to even or odd sites ${\bm r},{\bm r}'\in\mathbb{L}^{\pm}$.  We then decompose the quadratic Hamiltonian Eq.~\eqref{H2} in pairs of even-even, even-odd, odd-even, and odd-odd sites. The first line of Eq.~\eqref{H2} can be written as 
	\begin{align*}
	\hat{\mathcal{A}}=&   \sum_{{\bm r},{\bm r}'\in\mathbb{L}}\sum_{\alpha,\beta>0} \hat{\gamma}_{{\bm r},\alpha}^{\dag} A_{{\bm r}\alpha,{\bm r}'\beta}\hat{\gamma}_{{\bm r}',\beta}\\
	&= \hat{\mathcal{A}}^{\mathrm{ee}}+\hat{\mathcal{A}}^{\mathrm{oo}}+\hat{\mathcal{A}}^{\mathrm{eo}}+\hat{\mathcal{A}}^{\mathrm{oe}}.
	\end{align*}
	
	In order to transform the various terms to ${\bm k}$-space, it is useful to notice the relations
	\begin{align}
	\sum_{{\bm r}\in\mathbb{L}^{\pm}}e^{i{\bm k}\cdot{\bm r}}=\frac{N}{2}\left[\delta_{{\bm k},{\bm 0}}\pm\delta_{{\bm k}+\boldsymbol{\pi},{\bm 0}}\right]
	\end{align}
	with $\boldsymbol{\pi}=(\pi,\pi)^T$. 
	Using this relation, for the Fourier transformation of the even-even and odd-odd terms we find
	\begin{align*}
	\hat{\mathcal{A}}^{\mathrm{ee}}=&\sum_{{\bm r}\in\mathbb{L}^+,{\bm r}'\in\mathbb{L}^+}\sum_{\alpha,\beta>0} \hat{\gamma}_{{\bm r},\alpha}^{\dag} \left[A^{(0)}_{\mathrm{e}\alpha,\mathrm{e}\beta}\delta_{{\bm r},{\bm r}'}+A^{\mathrm{LR}}_{\mathrm{e}\alpha,\mathrm{e}\beta}\right]\hat{\gamma}_{{\bm r}',\beta}  \\
	=&\frac{1}{4}\sum_{{\bm k}\in\mathbb{B}}\sum_{\alpha,\beta>0}\left[\hat{\gamma}_{{\bm k},\alpha}^{\dag}+\hat{\gamma}_{{\bm k}+\boldsymbol{\pi},\alpha}^{\dag}\right] A^{(0)}_{\mathrm{e}\alpha,\mathrm{e}\beta}\left[\hat{\gamma}_{{\bm k},\beta}+\hat{\gamma}_{{\bm k}+\boldsymbol{\pi},\beta}\right]\\
	&+\frac{N}{4}\sum_{\alpha,\beta>0}\left[\hat{\gamma}^{\dag}_{{\bm 0},\alpha}+\hat{\gamma}^{\dag}_{\boldsymbol{\pi},\alpha}\right] A^{\mathrm{LR}}_{\mathrm{e}\alpha,\mathrm{e}\beta}\left[\hat{\gamma}_{{\bm 0},\beta}+\hat{\gamma}_{\boldsymbol{\pi},\beta}\right]\\
	=&\frac{1}{4}\sum_{{\bm k}\in\mathbb{B}}\sum_{\alpha,\beta>0}\left[\hat{\gamma}^{\dag}_{{\bm k},\alpha}+\hat{\gamma}^{\dag}_{{\bm k}+\boldsymbol{\pi},\alpha}\right]A_{\bm k,\alpha\beta}^{\mathrm{e}}\left[\hat{\gamma}_{{\bm k},\beta}+\hat{\gamma}_{{\bm k}+\boldsymbol{\pi},\beta}\right],\\
	\hat{\mathcal{A}}^{\mathrm{oo}}=&\sum_{{\bm r}\in\mathbb{L}^-,{\bm r}'\in\mathbb{L}^-}\sum_{\alpha,\beta>0} \hat{\gamma}_{{\bm r},\alpha}^{\dag} \left[A^{(0)}_{\mathrm{o}\alpha,\mathrm{o}\beta}\delta_{{\bm r},{\bm r}'}+A^{\mathrm{LR}}_{\mathrm{o}\alpha,\mathrm{o}\beta}\right]\hat{\gamma}_{{\bm r}',\beta}  \\
	=&\frac{1}{4}\sum_{{\bm k}\in\mathbb{B}}\sum_{\alpha,\beta>0}\left[\hat{\gamma}_{{\bm k},\alpha}^{\dag}-\hat{\gamma}_{{\bm k}+\boldsymbol{\pi},\alpha}^{\dag}\right] A^{(0)}_{\mathrm{o}\alpha,\mathrm{o}\beta}\left[\hat{\gamma}_{{\bm k},\beta}-\hat{\gamma}_{{\bm k}+\boldsymbol{\pi},\beta}\right]\\
	&+\frac{N}{4}\sum_{\alpha,\beta>0}\left[\hat{\gamma}^{\dag}_{{\bm 0},\alpha}-\hat{\gamma}^{\dag}_{\boldsymbol{\pi},\alpha}\right] A^{\mathrm{LR}}_{\mathrm{o}\alpha,\mathrm{o}\beta}\left[\hat{\gamma}_{{\bm 0},\beta}-\hat{\gamma}_{\boldsymbol{\pi},\beta}\right]\\
	=&\frac{1}{4}\sum_{{\bm k}\in\mathbb{B}}\sum_{\alpha,\beta>0}\left[\hat{\gamma}^{\dag}_{{\bm k},\alpha}-\hat{\gamma}^{\dag}_{{\bm k}+\boldsymbol{\pi},\alpha}\right]A_{\bm k,\alpha\beta}^{\mathrm{o}}\left[\hat{\gamma}_{{\bm k},\beta}-\hat{\gamma}_{{\bm k}+\boldsymbol{\pi},\beta}\right].
	\end{align*}
	
	while for the even-odd and odd-even terms we find
	\begin{align*}
	\hat{\mathcal{A}}^{\mathrm{eo}}=&\sum_{{\bm r}\in\mathbb{L}^+,{\bm r}'\in\mathbb{L}^-}\sum_{\alpha,\beta>0} \hat{\gamma}_{{\bm r},\alpha}^{\dag} \left[A^{\mathrm{hop}}_{\mathrm{e}\alpha,\mathrm{o}\beta}\delta_{{\bm r}',\mathcal{N}({\bm r})}+A^{\mathrm{LR}}_{\mathrm{e}\alpha,\mathrm{o}\beta}\right]\hat{\gamma}_{{\bm r}',\beta}\\
	=&\frac{1}{4}\sum_{{\bm k}\in\mathbb{B}}\sum_{\alpha,\beta>0}\left[\hat{\gamma}_{{\bm k},\alpha}^{\dag}+\hat{\gamma}_{{\bm k}+\boldsymbol{\pi},\alpha}^{\dag}\right]A^{\mathrm{hop}}_{\mathrm{e}\alpha,\mathrm{o}\beta}f_{\bm k}\left[\hat{\gamma}_{{\bm k},\beta}-\hat{\gamma}_{{\bm k}+\boldsymbol{\pi},\beta}\right]\\
	&+\frac{N}{4}\sum_{\alpha,\beta>0}\left[\hat{\gamma}^{\dag}_{{\bm 0},\alpha}+\hat{\gamma}^{\dag}_{\boldsymbol{\pi},\alpha}\right] A^{\mathrm{LR}}_{\mathrm{e}\alpha,\mathrm{o}\beta}\left[\hat{\gamma}_{{\bm 0},\beta}-\hat{\gamma}_{\boldsymbol{\pi},\beta}\right],\\
	=&\frac{1}{4}\sum_{{\bm k}\in\mathbb{B}}\sum_{\alpha,\beta>0}\left[\hat{\gamma}^{\dag}_{{\bm k},\alpha}+\hat{\gamma}^{\dag}_{{\bm k}+\boldsymbol{\pi},\alpha}\right]A_{\bm k,\alpha\beta}^{\mathrm{eo}}\left[\hat{\gamma}_{{\bm k},\beta}-\hat{\gamma}_{{\bm k}+\boldsymbol{\pi},\beta}\right],\\
	\hat{\mathcal{A}}^{\mathrm{oe}}=&\sum_{{\bm r}\in\mathbb{L}^-,{\bm r}'\in\mathbb{L}^+}\sum_{\alpha,\beta>0} \hat{\gamma}_{{\bm r},\alpha}^{\dag} \left[A^{\mathrm{hop}}_{\mathrm{o}\alpha,\mathrm{e}\beta}\delta_{{\bm r}',\mathcal{N}({\bm r})}+A^{\mathrm{LR}}_{\mathrm{o}\alpha,\mathrm{e}\beta}\right]\hat{\gamma}_{{\bm r}',\beta}\\
	=&\frac{1}{4}\sum_{{\bm k}\in\mathbb{B}}\sum_{\alpha,\beta>0}\left[\hat{\gamma}_{{\bm k},\alpha}^{\dag}-\hat{\gamma}_{{\bm k}+\boldsymbol{\pi},\alpha}^{\dag}\right]A^{\mathrm{hop}}_{\mathrm{o}\alpha,\mathrm{e}\beta}f_{\bm k}\left[\hat{\gamma}_{{\bm k},\beta}+\hat{\gamma}_{{\bm k}+\boldsymbol{\pi},\beta}\right]\\
	&+\frac{N}{4}\sum_{\alpha,\beta>0}\left[\hat{\gamma}^{\dag}_{{\bm 0},\alpha}-\hat{\gamma}^{\dag}_{\boldsymbol{\pi},\alpha}\right] A^{\mathrm{LR}}_{\mathrm{o}\alpha,\mathrm{e}\beta}\left[\hat{\gamma}_{{\bm 0},\beta}+\hat{\gamma}_{\boldsymbol{\pi},\beta}\right]\\
	=&\frac{1}{4}\sum_{{\bm k}\in\mathbb{B}}\sum_{\alpha,\beta>0}\left[\hat{\gamma}^{\dag}_{{\bm k},\alpha}-\hat{\gamma}^{\dag}_{{\bm k}+\boldsymbol{\pi},\alpha}\right]A_{\bm k,\alpha\beta}^{\mathrm{oe}}\left[\hat{\gamma}_{{\bm k},\beta}+\hat{\gamma}_{{\bm k}-\boldsymbol{\pi},\beta}\right].
	\end{align*}
	
	In these equations we have introduced the notation
	\begin{align}
	A_{\bm k,\alpha\beta}^{\mathrm{e}}=&(E^{\mathrm{MF}}_{\mathrm{e},\alpha}-E^{\mathrm{MF}}_{\mathrm{e},0})\delta_{\alpha,\beta}-U_{\mathrm{LR}} n_{\mathrm{e},\alpha0} n_{\mathrm{e},0\beta}[\delta_{{\bm k},0}+\delta_{{\bm k},\boldsymbol{\pi}}]\nonumber\\
	A_{\bm k,\alpha\beta}^{\mathrm{o}}=&(E^{\mathrm{MF}}_{\mathrm{o},\alpha}-E^{\mathrm{MF}}_{\mathrm{o},0})\delta_{\alpha,\beta}-U_{\mathrm{LR}}n_{\mathrm{o},\alpha0} n_{\mathrm{o},0\beta}[\delta_{{\bm k},0}+\delta_{{\bm k},\boldsymbol{\pi}}]\nonumber\\
	A_{\bm k,\alpha\beta}^{\mathrm{eo}}=&U_{\mathrm{LR}}n_{\mathrm{e},\alpha0}n_{\mathrm{o},0\beta}[\delta_{{\bm k},0}-\delta_{{\bm k},\boldsymbol{\pi}}]\nonumber\\
	&-tf_{\bm k}[b^\dagger_{\mathrm{e},\alpha0}b_{\mathrm{o},0\beta}+b^{\dagger}_{\mathrm{o},0\beta}b_{\mathrm{e},\alpha0}]\nonumber\\
	A_{\bm k,\alpha\beta}^{\mathrm{oe}}=&U_{\mathrm{LR}}n_{\mathrm{o},\alpha0}n_{\mathrm{e},0\beta}[\delta_{{\bm k},0}-\delta_{{\bm k},\boldsymbol{\pi}}]\nonumber\\
	&-tf_{\bm k}[b^\dagger_{\mathrm{o},\alpha0}b_{\mathrm{e},0\beta}+b^\dagger_{\mathrm{e},0\beta}b_{\mathrm{o},\alpha0}]\label{Ak} \ , 
	\end{align}
	where we introduced the expression $f_{\bm k}=2\left(\cos(k_x)+\cos(k_y)\right)$ in the hopping term ($k_x={\bm k}\cdot{\bm e}_x$, $k_y={\bm k}\cdot{\bm e}_y$). 
	Adding all those terms results in
	\begin{align}
	\hat{\mathcal{A}}=\frac{1}{2}\sum_{{\bm k}\in\mathbb{B}}\begin{pmatrix}
	\hat{\boldsymbol{\gamma}}^\dag_{\bm k}&\hat{\boldsymbol{\gamma}}^\dag_{\bm k+\boldsymbol{\pi}}
	\end{pmatrix}
	{\bm A}_{\bm k}
	\begin{pmatrix}
	[\hat{\boldsymbol{\gamma}}_{\bm k}]^T\\
	[\hat{\boldsymbol{\gamma}}_{\bm k+\boldsymbol{\pi}}]^T
	\end{pmatrix}.
	\end{align}
	with $\hat{\boldsymbol{\gamma}}_{\bm k}=(\hat{{\gamma}}_{\bm k,\alpha=1},\hat{{\gamma}}_{{\bm k},2},\dots)$ and
	\begin{align}
	{\bm A}_{\bm k}=&\begin{pmatrix}
	\frac{{\bm A}_{\bm k}^{\mathrm{e}}+{\bm A}_{\bm k}^{\mathrm{o}}}{2}+\frac{{\bm A}_{\bm k}^{\mathrm{eo}}+{\bm A}_{\bm k}^{\mathrm{oe}}}{2}&\frac{{\bm A}_{\bm k}^{\mathrm{e}}-{\bm A}_{\bm k}^{\mathrm{o}}}{2}-\frac{{\bm A}_{\bm k}^{\mathrm{eo}}-{\bm A}_{\bm k}^{\mathrm{oe}}}{2}\\
	\frac{{\bm A}_{\bm k}^{\mathrm{e}}-{\bm A}_{\bm k}^{\mathrm{o}}}{2}+\frac{{\bm A}_{\bm k}^{\mathrm{eo}}-{\bm A}_{\bm k}^{\mathrm{oe}}}{2}&\frac{{\bm A}_{\bm k}^{\mathrm{e}}+{\bm A}_{\bm k}^{\mathrm{o}}}{2}-\frac{{\bm A}_{\bm k}^{\mathrm{eo}}+{\bm A}_{\bm k}^{\mathrm{oe}}}{2}
	\end{pmatrix}.\label{Ak1}
	\end{align}
	Notice that here we have interpreted the results in Eq.~\eqref{Ak} as matrices with entries indexed by $(\alpha ,\beta)$. Whenever we numerically calculate these matrices we have to implement a cut-off that we introduced in the main text as $n_{\mathrm{max}}$. This cut-off is chosen to be  sufficiently large such that the results are insensitive to it. As a consequence of this cut-off the matrix ${\bm A}_{\bm k}$ is a square matrix with size $2n_{\mathrm{max}}\times2n_{\mathrm{max}}$.

	The first term of the second line of Eq.~\eqref{H2} can be written as 
	\begin{align*}
	\hat{\mathcal{B}}=&   \sum_{{\bm r},{\bm r}'\in\mathbb{L}}\sum_{\alpha,\beta>0} \hat{\gamma}_{{\bm r},\alpha} B_{{\bm r}\alpha,{\bm r}'\beta}\hat{\gamma}_{{\bm r}',\beta}\\
	&= \hat{\mathcal{B}}^{\mathrm{ee}}+\hat{\mathcal{B}}^{\mathrm{oo}}+\hat{\mathcal{B}}^{\mathrm{eo}}+\hat{\mathcal{B}}^{\mathrm{oe}}.
	\end{align*}
	In the same fashion as before we can calculate
	\begin{align*}
	\hat{\mathcal{B}}^{\mathrm{ee}}=&\sum_{{\bm r}\in\mathbb{L}^+,{\bm r}'\in\mathbb{L}^+}\sum_{\alpha,\beta>0} \hat{\gamma}_{{\bm r},\alpha} B^{\mathrm{LR}}_{\mathrm{e}\alpha,\mathrm{e}\beta}\hat{\gamma}_{{\bm r}',\beta} \\
	= &\frac{N}{4}\sum_{\alpha,\beta>0}\left[\hat{\gamma}_{{\bm 0},\alpha}+\hat{\gamma}_{\boldsymbol{\pi},\alpha}\right] B^{\mathrm{LR}}_{\mathrm{e}\alpha,\mathrm{e}\beta}\left[\hat{\gamma}_{{\bm 0},\beta}+\hat{\gamma}_{\boldsymbol{\pi},\beta}\right]\\
	=&\frac{1}{4}\sum_{{\bm k}\in\mathbb{B}}\sum_{\alpha,\beta>0}\left[\hat{\gamma}_{-{\bm k},\alpha}+\hat{\gamma}_{-{\bm k}+\boldsymbol{\pi},\alpha}\right]B_{\bm k,\alpha\beta}^{\mathrm{e}}\left[\hat{\gamma}_{{\bm k},\beta}+\hat{\gamma}_{{\bm k}+\boldsymbol{\pi},\beta}\right],\\
	\hat{\mathcal{B}}^{\mathrm{oo}}=&\sum_{{\bm r}\in\mathbb{L}^-,{\bm r}'\in\mathbb{L}^-}\sum_{\alpha,\beta>0} \hat{\gamma}_{{\bm r},\alpha} B^{\mathrm{LR}}_{\mathrm{o}\alpha,\mathrm{o}\beta}\hat{\gamma}_{{\bm r}',\beta}\\
	=&\frac{N}{4}\sum_{\alpha,\beta>0}\left[\hat{\gamma}_{{\bm 0},\alpha}-\hat{\gamma}_{\boldsymbol{\pi},\alpha}\right] B^{\mathrm{LR}}_{\mathrm{o}\alpha,\mathrm{o}\beta}\left[\hat{\gamma}_{{\bm 0},\beta}-\hat{\gamma}_{\boldsymbol{\pi},\beta}\right]\\
	=&\frac{1}{4}\sum_{{\bm k}\in\mathbb{B}}\sum_{\alpha,\beta>0}\left[\hat{\gamma}_{-{\bm k},\alpha}-\hat{\gamma}_{-{\bm k}+\boldsymbol{\pi},\alpha}\right]B_{\bm k,\alpha\beta}^{\mathrm{o}}\left[\hat{\gamma}_{{\bm k},\beta}-\hat{\gamma}_{{\bm k}+\boldsymbol{\pi},\beta}\right].
	\end{align*}
	Now we calculate
	\begin{align*}
	\hat{\mathcal{B}}^{\mathrm{eo}}=&\sum_{{\bm r}\in\mathbb{L}^+,{\bm r}'\in\mathbb{L}^-}\sum_{\alpha,\beta>0} \hat{\gamma}_{{\bm r},\alpha} \left[B^{\mathrm{hop}}_{\mathrm{e}\alpha,\mathrm{o}\beta}\delta_{{\bm r}',\mathcal{N}({\bm r})}+B^{\mathrm{LR}}_{\mathrm{e}\alpha,\mathrm{o}\beta}\right]\hat{\gamma}_{{\bm r}',\beta}\\
	=&\frac{1}{4}\sum_{{\bm k}\in\mathbb{B}}\sum_{\alpha,\beta>0}\left[\hat{\gamma}_{-{\bm k},\alpha}+\hat{\gamma}_{-{\bm k}+\boldsymbol{\pi},\alpha}\right]B^{\mathrm{hop}}_{\mathrm{e}\alpha,\mathrm{o}\beta}f_{\bm k}\left[\hat{\gamma}_{{\bm k},\beta}-\hat{\gamma}_{{\bm k}+\boldsymbol{\pi},\beta}\right]\\
	&+\frac{N}{4}\sum_{\alpha,\beta>0}\left[\hat{\gamma}_{{\bm 0},\alpha}+\hat{\gamma}_{\boldsymbol{\pi},\alpha}\right] B^{\mathrm{LR}}_{\mathrm{e}\alpha,\mathrm{o}\beta}\left[\hat{\gamma}_{{\bm 0},\beta}-\hat{\gamma}_{\boldsymbol{\pi},\beta}\right],\\
	=&\frac{1}{4}\sum_{{\bm k}\in\mathbb{B}}\sum_{\alpha,\beta>0}\left[\hat{\gamma}_{-{\bm k},\alpha}+\hat{\gamma}_{-{\bm k}+\boldsymbol{\pi},\alpha}\right]B_{\bm k,\alpha\beta}^{\mathrm{eo}}\left[\hat{\gamma}_{{\bm k},\beta}-\hat{\gamma}_{{\bm k}+\boldsymbol{\pi},\beta}\right],\\
	\hat{\mathcal{B}}^{\mathrm{oe}}=&\sum_{{\bm r}\in\mathbb{L}^-,{\bm r}'\in\mathbb{L}^+}\sum_{\alpha,\beta>0} \hat{\gamma}_{{\bm r},\alpha} \left[B^{\mathrm{hop}}_{\mathrm{o}\alpha,\mathrm{e}\beta}\delta_{{\bm r}',\mathcal{N}({\bm r})}+B^{\mathrm{LR}}_{\mathrm{o}\alpha,\mathrm{e}\beta}\right]\hat{\gamma}_{{\bm r}',\beta}\\
	=&\frac{1}{4}\sum_{{\bm k}\in\mathbb{B}}\sum_{\alpha,\beta>0}\left[\hat{\gamma}_{-{\bm k},\alpha}-\hat{\gamma}_{-{\bm k}+\boldsymbol{\pi},\alpha}\right]B^{\mathrm{hop}}_{\mathrm{o}\alpha,\mathrm{e}\beta}f_{\bm k}\left[\hat{\gamma}_{{\bm k},\beta}+\hat{\gamma}_{{\bm k}+\boldsymbol{\pi},\beta}\right]\\
	&+\frac{N}{4}\sum_{\alpha,\beta>0}\left[\hat{\gamma}_{{\bm 0},\alpha}-\hat{\gamma}_{\boldsymbol{\pi},\alpha}\right] B^{\mathrm{LR}}_{\mathrm{o}\alpha,\mathrm{e}\beta}\left[\hat{\gamma}_{{\bm 0},\beta}+\hat{\gamma}_{\boldsymbol{\pi},\beta}\right]\\
	=&\frac{1}{4}\sum_{{\bm k}\in\mathbb{B}}\sum_{\alpha,\beta>0}\left[\hat{\gamma}_{-{\bm k},\alpha}-\hat{\gamma}_{-{\bm k}+\boldsymbol{\pi},\alpha}\right]B_{\bm k,\alpha\beta}^{\mathrm{oe}}\left[\hat{\gamma}_{{\bm k},\beta}+\hat{\gamma}_{{\bm k}+\boldsymbol{\pi},\beta}\right].
	\end{align*}
	Here we have used the definitions
	\begin{align}
	B_{\bm k,\alpha\beta}^{\mathrm{e}}=&-U_{\mathrm{LR}} n_{\mathrm{e},0\alpha}n_{\mathrm{e},0\beta}[\delta_{{\bm k},0}+\delta_{{\bm k},\boldsymbol{\pi}}]\nonumber\\
	B_{\bm k,\alpha\beta}^{\mathrm{o}}=&-U_{\mathrm{LR}} n_{\mathrm{o},0\alpha}n_{\mathrm{o},0\beta}[\delta_{{\bm k},0}+\delta_{{\bm k},\boldsymbol{\pi}}]\nonumber\\
	B_{\bm k,\alpha\beta}^{\mathrm{eo}}=&U_{\mathrm{LR}} n_{\mathrm{e},0\alpha}n_{\mathrm{o},0\beta}[\delta_{{\bm k},0}-\delta_{{\bm k},\boldsymbol{\pi}}]\nonumber\\
	&-tf_{\bm k}[b^\dagger_{\mathrm{e},0\alpha}b_{\mathrm{o},0\beta}+b^{\dagger}_{\mathrm{o},0\beta}b_{\mathrm{e},0\alpha}]\nonumber\\
	B_{\bm k,\alpha\beta}^{\mathrm{oe}}=&U_{\mathrm{LR}} n_{\mathrm{o},0\alpha}n_{\mathrm{e},0\beta}[\delta_{{\bm k},0}-\delta_{{\bm k},\boldsymbol{\pi}}]\nonumber\\
	&-tf_{\bm k}[b^\dagger_{\mathrm{o},0\alpha}b_{\mathrm{e},0\beta}+b^\dagger_{\mathrm{e},0\beta}b_{\mathrm{o},0\alpha}].\label{Bk}
	\end{align}
	Now we can add all terms to obtain
	\begin{align}
	\hat{\mathcal{B}}=\frac{1}{2}\sum_{{\bm k}\in\mathbb{B}}\begin{pmatrix}
	\hat{\boldsymbol{\gamma}}_{-\bm k}&\hat{\boldsymbol{\gamma}}_{-\bm k+\boldsymbol{\pi}}
	\end{pmatrix}
	{\bm B}_{\bm k}
	\begin{pmatrix}
	[\hat{\boldsymbol{\gamma}}_{\bm k}]^T\\
	[\hat{\boldsymbol{\gamma}}_{\bm k+\boldsymbol{\pi}}]^T
	\end{pmatrix},
	\end{align}
	where we used
	\begin{align}
	{\bm B}_{\bm k}=&\begin{pmatrix}
	\frac{{\bm B}_{\bm k}^{\mathrm{e}}+{\bm B}_{\bm k}^{\mathrm{o}}}{2}+\frac{{\bm B}_{\bm k}^{\mathrm{eo}}+{\bm B}_{\bm k}^{\mathrm{oe}}}{2}&\frac{{\bm B}_{\bm k}^{\mathrm{e}}-{\bm B}_{\bm k}^{\mathrm{o}}}{2}-\frac{{\bm B}_{\bm k}^{\mathrm{eo}}-{\bm B}_{\bm k}^{\mathrm{oe}}}{2}\\
	\frac{{\bm B}_{\bm k}^{\mathrm{e}}-{\bm B}_{\bm k}^{\mathrm{o}}}{2}+\frac{{\bm B}_{\bm k}^{\mathrm{eo}}-{\bm B}_{\bm k}^{\mathrm{oe}}}{2}&\frac{{\bm B}_{\bm k}^{\mathrm{e}}+{\bm B}_{\bm k}^{\mathrm{o}}}{2}-\frac{{\bm B}_{\bm k}^{\mathrm{eo}}+{\bm B}_{\bm k}^{\mathrm{oe}}}{2}
	\end{pmatrix}.\label{Bk1}
	\end{align}
	With the previously mentioned cutoff $n_{\mathrm{max}}$, ${\bm B}_{\bm k}$ is also a square matrix with size $2n_{\mathrm{max}}\times2n_{\mathrm{max}}$.
	Using the Hermitian conjugate we find
	\begin{align}
	\hat{\mathcal{B}}^{\dag}=\frac{1}{2}\sum_{{\bm k}\in\mathbb{B}}\begin{pmatrix}
	\hat{\boldsymbol{\gamma}}_{\bm k}^{\dag}&\hat{\boldsymbol{\gamma}}_{\bm k+\boldsymbol{\pi}}^\dag
	\end{pmatrix}
	{\bm B}_{\bm k}
	\begin{pmatrix}
	[\hat{\boldsymbol{\gamma}}^{\dag}_{-\bm k}]^T\\
	[\hat{\boldsymbol{\gamma}}^{\dag}_{-\bm k+\boldsymbol{\pi}}]^T
	\end{pmatrix},
	\end{align}
	where we used ${\bm B}_{\bm k}^{\dag}={\bm B}_{\bm k}$. 
	
	Finally, using
	\begin{align}
	\hat{H}^{(2)}=\hat{\mathcal{A}}+\frac{\hat{\mathcal{B}}+\hat{\mathcal{B}}^{\dag}}{2},
	\end{align}
	we obtain the result
	\begin{eqnarray}\label{hfin}
	\hat{H}^{(2)}=\frac{1}{4}\sum_{{\bm k}\in\mathbb{B}} \hat{\boldsymbol{\Gamma}}_{\bm k}^{\dag}{\bm H}^{(2)}_{\bm k}\hat{\boldsymbol{\Gamma}}_{\bm k}^T,
	\end{eqnarray}
	with $\hat{\boldsymbol{\Gamma}}_{\bm k}=(\hat{\boldsymbol{\gamma}}_{\bm k},\hat{\boldsymbol{\gamma}}_{{\bm k}+\boldsymbol{\pi}}, \hat{\boldsymbol{\gamma}}^{\dag}_{-\bm k},\hat{\boldsymbol{\gamma}}^{\dag}_{-{\bm k}-\boldsymbol{\pi}})$ and $\boldsymbol{\pi}=(\pi,\pi)^T$. 
	The coupling matrix is given by 
	\begin{align}\label{H_keo}
	{\bm H}^{(2)}_{\bm k}=\begin{pmatrix}
	{\bm A}_{\bm k}&{\bm B}_{\bm k}\\
	{\bm B}_{\bm k}&{\bm A}_{\bm k}
	\end{pmatrix}.
	\end{align}
	Here, ${\bm H}^{(2)}_{\bm k}$ is a $4n_{\mathrm{max}}\times4n_{\mathrm{max}}$ square matrix.
	The diagonal Hamiltonian can be diagonalized by a Bogolyubov transformation \cite{blaizot-ripka} for every value of ${\bm k}$, defined by the matrix ${\bm T}_{\bm k}$ such that
	\begin{align}
	{\bm T}_{\bm k}^{\dag}{\bm H}^{(2)}_{\bm k}{\bm T}_{\bm k}={\bm \Omega}_{\bm k}
	\end{align}
	and
	\begin{align}
	{\bm \Omega}_{\bm k}=\begin{pmatrix}
	\boldsymbol{\omega}_{\bm k}&{\bm 0}_{2n_{\mathrm{max}}}\\
	{\bm 0}_{2n_{\mathrm{max}}}&\boldsymbol{\omega}_{\bm k}
	\end{pmatrix}
	\end{align}
	Here, $\boldsymbol{\omega}_{\bm k}=\mathrm{diag}(\omega_{{\bm k},1},\omega_{{\bm k},2}.\dots,\omega_{{\bm k},2n_{\mathrm{max}}})$ is a diagonal matrix containing the $2n_{\mathrm{max}}$ excitation energies corresponding to the wave vector pair ${\bm k}$ and ${\bm k}+\boldsymbol{\pi}$. The matrix ${\bm T}_{\bm k}$ is found by diagonalizing $\boldsymbol{\Upsilon}{\bm H}^{(2)}_{\bm k}$ where $\boldsymbol{\Upsilon}$ is the matrix 
	\begin{align}
	\boldsymbol{\Upsilon} =\begin{pmatrix}
	{\bm 1}_{2n_{\mathrm{max}}}&{\bm 0}_{2n_{\mathrm{max}}}\\
	{\bm 0}_{2n_{\mathrm{max}}}&{-\bm 1}_{2n_{\mathrm{max}}}
	\end{pmatrix}  
	\end{align}and imposing the normalization ${\bm T}_{\bm k}^{\dag}\boldsymbol{\Upsilon}{\bm T}_{\bm k}=\boldsymbol{\Upsilon}$.  We have used the notation where ${\bm 1}_{M}$ is the $M\times M$ identity matrix and ${\bm 0}_{M}$ is the $M\times M$ matrix with only zeros. With the inverse transformation 
	\begin{align}
	{\bm T}_{\bm k} \hat{{\bm \Delta}}_{\bm k}^T=\hat{\boldsymbol{\Gamma}}_{\bm k}^T\label{Deltatransform}
	\end{align}
	we can then find the bosonic eigenmodes $\hat{\bm \Delta}_{\bm k}$ and the excitation energies given by $\boldsymbol{\omega}_{\bm k}$.
	
	\subsection{From the complete to the reduced Brillouin zone} \label{kreprent}
	
	In this section we discuss the modifications of the Brillouin zone due to the staggered long-range interactions.
	
	Equation~\eqref{hfin} shows that every ${\bm k}$ mode is beside the usual coupling to $-{\bm k}$ also coupled to the shifted vectors ${\bm k}+\boldsymbol{\pi}$ and $-{\bm k}-\boldsymbol{\pi}$. The origin of this coupling is the even-odd imbalance. In fact, when we assume that there is no even-odd imbalance we find that ${\bm A}_{\bm k}^\mathrm{e}={\bm A}_{\bm k}^\mathrm{o}$ and ${\bm A}_{\bm k}^\mathrm{eo}={\bm A}_{\bm k}^\mathrm{oe}$ and the same for the matrix ${\bm B}_{\bm k}$: therefore the off-diagonal blocks in Eqs.~\eqref{Ak1} and \eqref{Bk1} vanish and the modes ${\bm k}$ and ${\bm k}+\boldsymbol{\pi}$ are decoupled. Without even-odd imbalance, in the SF and MI phases, we can therefore find the eigenmodes and eigenenergies for ${\bm k}$ and ${\bm k}+\boldsymbol{\pi}$ separately, while this is not possible in presence of even-odd imbalance, \textit{i.e.}, in the SS and CDW phases.
	
\begin{figure}[h!!!]
		{\includegraphics[width=0.9\linewidth]{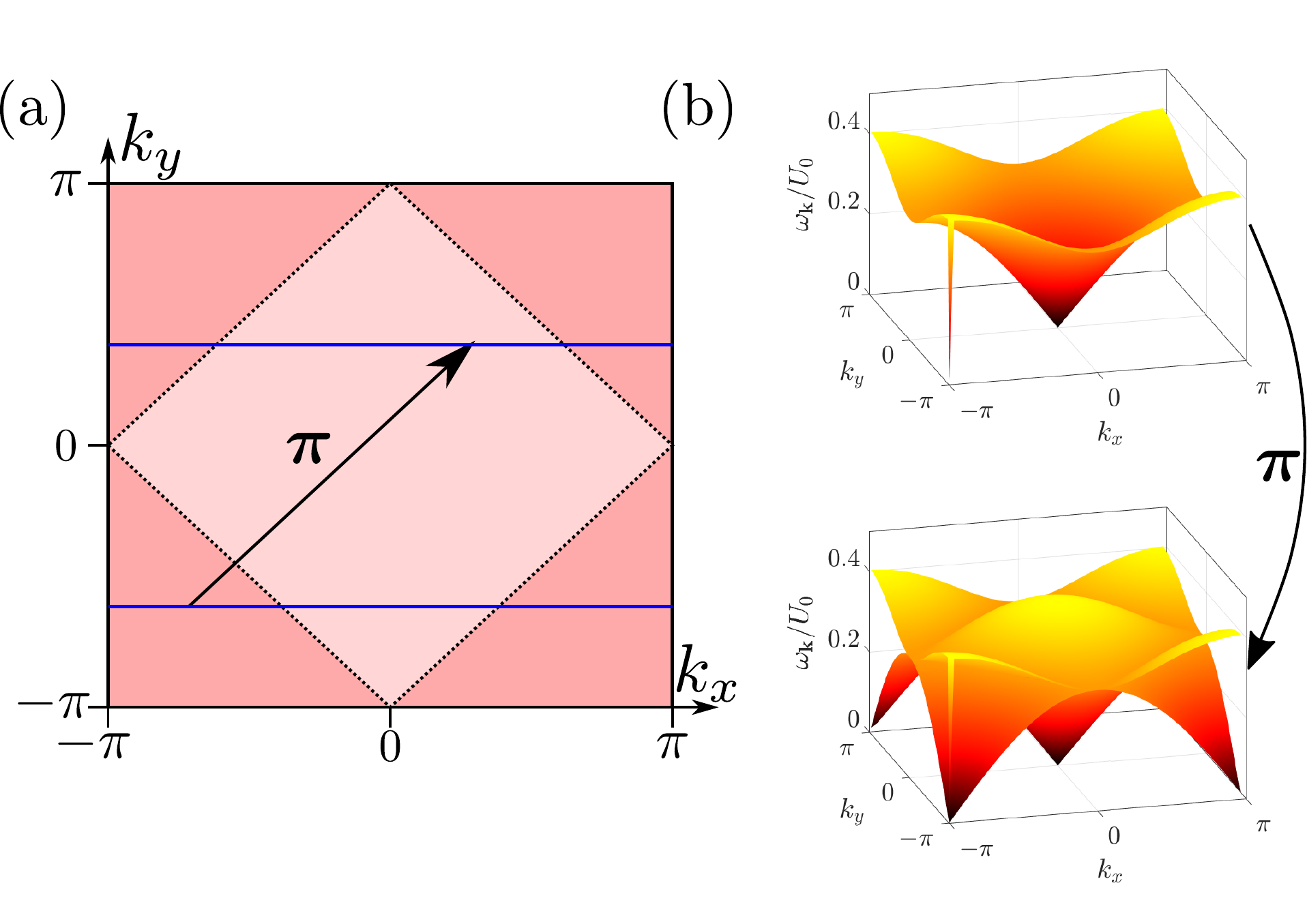}}
		\caption{(a) Visualization of the reduction of the Brillouin zone when a coupling between the wavevectors ${\bm k}$ to ${\bm k}+\boldsymbol{\pi}$ is induced. The corners (dark pink) are folded inwards (light pink). Wavevectors along the blue lines couple. (b) Representation of the same energy spectrum in the SF phase. Up: the Brillouin zone is not folded and we observe the lowest band. Down: After folding every energy is counted twice and there is a second band that is a copy of the lowest band. This band is displaced by $\boldsymbol{\pi}$. The data has been produced using $L=60$, $\mu=-0.052U_0$, $U_\mathrm{LR}=0.3U_0$, and $4t=0.2505U_0$.} \label{FigS:1}
	\end{figure}

	This can in fact be seen as a folding of the Brillouin zone  onto a reduced one (corresponding to the lattice with a two-site basis), as depicted in Fig.~S\ref{FigS:1}(a). For the sake of illustration, Fig.~S\ref{FigS:1}(b) shows the lowest branch of the dispersion relation of quasiparticle excitations in the SF phase (close to the SS/SF transition), and its folded version. Upon folding, the roton mode at $\pi,\pi$ (and equivalent vectors) is mapped onto a soft mode at $(0,0)$ (not visible in the figure). When entering into the SS phase, a gap opens along the edges of the reduced Brillouin zone, as we will further discuss in Sec.~\ref{EE} (see Fig.~\ref{FigS:4}).

	\subsection{Entanglement entropy and Entanglement spectrum} 
	In this section we discuss how we calculate the entanglement entropy and the entanglement spectrum.
	
	As pointed out in the main text, since we describe the entanglement in an effective quadratic system we can use the correlation matrix
	\begin{align}\label{crmat}
	\mathbf{C}_\mathbb{A} =&\sum_{{\bm r},{\bm r}'\in\mathbb{A}}\mathbf{C}_{{\bm r},{\bm r}'}|{\bm r}\rangle\langle{\bm r}'|,
	\end{align}
	with
	$\mathbf{C}_{{\bm r},{\bm r}'}=\langle (\hat{\boldsymbol{\gamma}}_{\bm r},\hat{\boldsymbol{\gamma}}^{\dag}_{\bm r})^T(\hat{\boldsymbol{\gamma}}^{\dag}_{{\bm r}'},\hat{\boldsymbol{\gamma}}_{{\bm r}'})\rangle$
	with  $\hat{\boldsymbol{\gamma}}_{\bm r}=(\hat{{\gamma}}_{\bm r,\alpha=1},\hat{{\gamma}}_{{\bm r},2},\dots,\hat{{\gamma}}_{{\bm r},n_{\mathrm{max}}})$. Here, $n_{\mathrm{max}}$ is the previously mentioned cut-off for the local Fock space. The description of $\mathbf{C}_\mathbb{A}$ is valid for any subsystem $\mathbb{A}\subset\mathbb{L}$.
	
	The matrix ${\bm C}_{\bm k}$ can be calculated from the eigenmodes in ${\bm k}$ space using
	\begin{align}
	{\bm C}_{\bm k}={\bm T}_{\bm k}{\bm P}{\bm T}_{\bm k}^{\dag},  
	\end{align}
	where we have used that all eigenmodes $\hat{\bm \Delta}_{\bm k}$ are in their vacuum state at zero temperature
	\begin{align}
	\left\langle\hat{\bm \Delta}^T_{\bm k}\hat{\bm \Delta}^{\dag}_{\bm k}\right\rangle ={\bm P}:= \begin{pmatrix}
	{\bm 1}&{\bm 0}\\
	{\bm 0}&{\bm 0}
	\end{pmatrix},
	\end{align}
	and ${\bm T}_{\bm k}\hat{\bm \Delta}_{\bm k}^T=\hat{\boldsymbol{\Gamma}}_{\bm k}^T$ is the Bogolyubov transformation that transforms from $\hat{\boldsymbol{\Gamma}}_{\bm k}$ to the eigenmodes $\hat{\bm \Delta}_{\bm k}$  [see Eq.~\eqref{Deltatransform}].

	In order to calculate the entanglement entropy from this correlation matrix we first need to find a Bogolyubov transformation ${\bm T}_{\mathbb A}$ such that 
	\begin{align}
	{\bm T}_{\mathbb A}^\dag{\bm C}_{\mathbb A}{\bm T}_\mathbb{A}=\begin{pmatrix}
	{\bm 1}_{\|\mathbb{A}\|\cdot n_{\mathrm{max}}}+{\bm n}~~&{\bm 0}_{\|\mathbb{A}\|\cdot n_{\mathrm{max}}}\\
	{\bm 0}_{\|\mathbb{A}\|\cdot n_{\mathrm{max}}}~~&{\bm n}
	\end{pmatrix}
	\end{align}
	with ${\bm n}$ a diagonal matrix contaning $\|\mathbb{A}\|\cdot n_{\mathrm{max}}$ entries $n^{(j)}$ on the diagonal. Here, $\|\mathbb{A}\|$ is the number of sites in $\mathbb{A}$. The ${\bm T}_{\mathbb A}$ matrix is normalized as ${\bm T}_{\mathbb A}^\dag\boldsymbol{\Upsilon}_\mathbb{A}{\bm T}_{\mathbb A}=\boldsymbol{\Upsilon}_{\mathbb A}$ with 
	
	$$\boldsymbol{\Upsilon}_{\mathbb A}=\begin{pmatrix}
	{\bm 1}_{\|\mathbb{A}\|\cdot n_{\mathrm{max}}}&{\bm 0}_{\|\mathbb{A}\|\cdot n_{\mathrm{max}}}\\
	{\bm 0}_{\|\mathbb{A}\|\cdot n_{\mathrm{max}}}&{ -\bm 1}_{\|\mathbb{A}\|\cdot n_{\mathrm{max}}}
	\end{pmatrix}.$$
	
	The entanglement entropy is then given by \cite{Frerot:2016}
	\begin{align}
	S=\sum_{j=1}^{\|\mathbb{A}\|\cdot n_{\mathrm{max}}} \left[\{1+n^{(j)}\}\log\{1+n^{(j)}\}-n^{(j)}\log\{n^{(j)}\}\right]. \label{EEwithN}   
	\end{align}
	
	We will now study the entanglement for the specific choice of subsystem that we have used in the main text, namely $\mathbb{A}$ corresponds to the half torus with $L/2\times L$ sites.  $\mathbb{A}$ can be written as $\mathbb{A}=\mathbb{A}_x\times\mathbb{A}_y$ with $\mathbb{A}_x=\{1, 2,\dots,L/2\}$ and $\mathbb{A}_y=\{1, 2, \dots, L\}$. This subsystem is translationally invariant along the $y$-direction, allowing us to use the Fourier-transformed basis:
	\begin{align}
	|y\rangle=\frac{1}{\sqrt{L}}\sum_{k_y\in\mathbb{B}_{y}}e^{-ik_y y}|k_y\rangle\label{Fouriertrafo},
	\end{align}
	where we have defined $\mathbb{B}_{y}=\{\frac{2\pi}{L}m ,m=-L/2,-L/2+1\dots,L/2-1\}$.
	
	It is convenient to decompose the correlation matrix into four blocks
	\begin{align}
	\mathbf{C}_\mathbb{A} =\begin{pmatrix} \mathbf{C}_\mathbb{A}^{(11)} & \mathbf{C}_\mathbb{A}^{(12)} \\ \mathbf{C}_\mathbb{A}^{(21)} & \mathbf{C}_\mathbb{A}^{(22)} \end{pmatrix} 
	\end{align}
	defined as 
	\begin{align}
	{\bm C}_\mathbb{A}^{11}=&\sum_{{\bm r},{\bm r}'\in\mathbb{A}}\langle \hat{\boldsymbol{\gamma}}_{{\bm r}}^T\hat{\boldsymbol{\gamma}}_{{\bm r}'}^{\dag}\rangle|{\bm r}\rangle\langle{\bm r}'| \\
	{\bm C}^{12}_\mathbb{A}=&\sum_{{\bm r},{\bm r}'\in\mathbb{A}}\langle \hat{\boldsymbol{\gamma}}_{{\bm r}}^T\hat{\boldsymbol{\gamma}}_{{\bm r}'}\rangle|{\bm r}\rangle\langle{\bm r}'|\\
	{\bm C}^{21}_\mathbb{A}=&\sum_{{\bm r},{\bm r}'\in\mathbb{A}}\langle [\hat{\boldsymbol{\gamma}}_{{\bm r}}^{\dag}]^T\hat{\boldsymbol{\gamma}}_{{\bm r}'}^{\dag}\rangle|{\bm r}\rangle\langle{\bm r}'|\\		
	{\bm C}^{22}_\mathbb{A}=&\sum_{{\bm r},{\bm r}'\in\mathbb{A}}\langle[ \hat{\boldsymbol{\gamma}}^{\dag}_{{\bm r}}]^T\hat{\boldsymbol{\gamma}}_{{\bm r}'}\rangle|{\bm r}\rangle\langle{\bm r}'|
	\end{align} 
	
	Now, defining ${\bm r}=(x,y)^T$ and ${\bm r}'=(x',y')^T$, and applying the Fourier transform only over the $y$ components we obtain for ${\bm C}_\mathbb{A}^{11}$
	\begin{align*}
	{\bm C}_\mathbb{A}^{11}=\sum_{x,x'\in\mathbb{A}_x}\sum_{k_y,k_y'\in \mathbb{B}_y}\langle \hat{\boldsymbol{\gamma}}^T_{x,k_y}\hat{\boldsymbol{\gamma}}_{x',k_y'}^{\dag}\rangle|x,k_y\rangle\langle x',k_y'|.
	\end{align*}
	We can now use the fact that 
	\begin{align*}
	\langle \hat{\boldsymbol{\gamma}}^T_{x,k_y}\hat{\boldsymbol{\gamma}}_{x',k_y'}^{\dag}\rangle=\left[\delta_{k_y,k_y'}+\delta_{k_y,k_y'+\pi}\right]  \langle \hat{\boldsymbol{\gamma}}^T_{x,k_y}\hat{\boldsymbol{\gamma}}_{x',k_y'}^{\dag}\rangle
	\end{align*}
	so that, defining a $\mathbb{B}_y^{<0}=\left\{k_y\in\mathbb{B}_y|k_y<0\right\}$, we can write
	\begin{align*}
	{\bm C}_\mathbb{A}^{11}=&\sum_{x,x'\in\mathbb{A}_x}\sum_{k_y\in \mathbb{B}_y^{<0}} {\bm C}^{11}_{k_y}(x,x')\otimes |x,k_y\rangle\langle x',k_y| ,
	\end{align*}
	
	where for every $k_y\in\mathbb{B}_y^{<0}$ we have introduced the $2n_{\mathrm{max}}\times2n_{\mathrm{max}}$ matrix ${\bm C}^{11}_{k_y}(x,x')$ (in a $2 \times 2$ block form spanned by $|k_y\rangle$ and $|k_y+\pi\rangle$), which reads
	
	\begin{align*}
	{\bm C}^{11}_{k_y}(x,x')=\left\langle\begin{pmatrix}
	\hat{\boldsymbol{\gamma}}^T_{x,k_y}\\
	\hat{\boldsymbol{\gamma}}^T_{x,k_y+\pi}
	\end{pmatrix}\begin{pmatrix}
	\hat{\boldsymbol{\gamma}}^{\dag}_{x',k_y}&
	\hat{\boldsymbol{\gamma}}^{\dag}_{x',k_y+\pi}
	\end{pmatrix}\right\rangle.
	\end{align*}
	
	Performing a similar calculation for the other blocks of the correlation matrix, we find
	\begin{align*}
	{\bm C}^{12}_\mathbb{A}=&\sum_{x,x'\in\mathbb{A}_x}\sum_{k_y,k_y'\in \mathbb{B}_y}\langle \hat{\boldsymbol{\gamma}}_{x,k_y}^T\hat{\boldsymbol{\gamma}}_{x',-k_y'}\rangle|x,k_y\rangle\langle x',k_y'|\\
	=&\sum_{x,x'\in\mathbb{A}_x}\sum_{k_y\in \mathbb{B}_y^{<0}} {\bm C}^{12}_{k_y}(x,x')\otimes |x,k_y\rangle\langle x',k_y|,\\
	{\bm C}^{21}_\mathbb{A}=&\sum_{x,x'\in\mathbb{A}_x}\sum_{k_y,k_y'\in \mathbb{B}_y}\langle[ \hat{\boldsymbol{\gamma}}^{\dag}_{x,-k_y}]^T\hat{\boldsymbol{\gamma}}^{\dag}_{x',k_y'}\rangle|x,k_y\rangle\langle x',k_y'|\\
	=&\sum_{x,x'\in\mathbb{A}_x}\sum_{k_y\in \mathbb{B}_y^{<0}} {\bm C}^{21}_{k_y}(x,x')\otimes |x,k_y\rangle\langle x',k_y|,\\
	{\bm C}^{22}_\mathbb{A}=&\sum_{x,x'\in\mathbb{A}_x}\sum_{k_y,k_y'\in \mathbb{B}_y}\langle[ \hat{\boldsymbol{\gamma}}^{\dag}_{x,-k_y}]^T\hat{\boldsymbol{\gamma}}_{x',-k_y'}\rangle|x,k_y\rangle\langle x',k_y'|\\
	=&\sum_{x,x'\in\mathbb{A}_x}\sum_{k_y\in \mathbb{B}_y^{<0}} {\bm C}^{22}_{k_y}(x,x')\otimes |x,k_y\rangle\langle x',k_y|,
	\end{align*}
	with the matrices
	\begin{align*}
	{\bm C}^{12}_{k_y}(x,x')=&\left\langle\begin{pmatrix}
	\hat{\boldsymbol{\gamma}}_{x,k_y}^T\\
	\hat{\boldsymbol{\gamma}}_{x,k_y+\pi}^T
	\end{pmatrix}\begin{pmatrix}
	\hat{\boldsymbol{\gamma}}_{x',-k_y}&
	\hat{\boldsymbol{\gamma}}_{x',-k_y+\pi}
	\end{pmatrix}\right\rangle\\
	{\bm C}^{21}_{k_y}(x,x')=&\left\langle\begin{pmatrix}
	[\hat{\boldsymbol{\gamma}}^{\dag}_{x,-k_y}]^T\\
	[\hat{\boldsymbol{\gamma}}^{\dag}_{x,-k_y+\pi}]^T
	\end{pmatrix}\begin{pmatrix}
	\hat{\boldsymbol{\gamma}}^{\dag}_{x',k_y}&
	\hat{\boldsymbol{\gamma}}^{\dag}_{x',k_y+\pi}
	\end{pmatrix}\right\rangle\\
	{\bm C}^{22}_{k_y}(x,x')=&\left\langle\begin{pmatrix}
	[\hat{\boldsymbol{\gamma}}^{\dag}_{x,-k_y}]^T\\
	[\hat{\boldsymbol{\gamma}}^{\dag}_{x,-k_y+\pi}]^T
	\end{pmatrix}\begin{pmatrix}
	\hat{\boldsymbol{\gamma}}_{x',-k_y}&
	\hat{\boldsymbol{\gamma}}_{x',-k_y+\pi}
	\end{pmatrix}\right\rangle.
	\end{align*}
	This shows that the correlation matrix is block diagonal with
	\begin{align}
	{\bm C}_\mathbb{A}=\sum_{k_y\in\mathbb{B}^{<0}_y}{\bm C}_{k_y}|k_y\rangle\langle k_y|\label{BlockdiagonalC}
	\end{align}
	and
	\begin{align*}
	{\bm C}_{k_y}=\sum_{x,x'\in\mathbb{A}_x}  {\bm C}_{k_y}(x,x')|x\rangle\langle x'|  .
	\end{align*}
	Here, we have used the notation
	\begin{align*}
	{\bm C}_{k_y}=\begin{pmatrix}
	{\bm C}^{11}_{k_y}&{\bm C}^{12}_{k_y}\\
	{\bm C}^{21}_{k_y}&{\bm C}^{22}_{k_y}
	\end{pmatrix}.
	\end{align*}

	Since Eq.~\eqref{BlockdiagonalC} shows that the correlation matrix is block-diagonal we can find a Bogolyubov transformations ${\bm T}_{k_y}$ that diagonalizes ${\bm C}_{k_y}$ for each $k_y$. As a consequence for every $k_y\in\mathbb{B}_y^{<0}$ we find a diagonal matrix ${\bm n}_{k_y}$ 
	such that
	\begin{align}
	{\bm n}=\sum_{k_y\in\mathbb{B}_y^{<0}}{\bm n}_{k_y}|k_y\rangle\langle k_y|
	\end{align}
	and we can use Eq.~\eqref{EEwithN} to calculate the entanglement entropy. 
	
	Moreover  we can calculate the entanglement spectrum $\boldsymbol{\lambda}_{k_y}$ from the diagonal matrix ${\bm n}_{k_y}$. Those two quantities are related element by element via the equation
	\begin{align}
	n^{(i)}_{k_y}=\frac{1}{e^{\lambda^{(i)}_{k_y}}-1}.\label{ES}
	\end{align}

	\section{Supplemental numerical results of the Entanglement Entropy}
	\label{EE}
	
	In this section we show additional numerical results of the entanglement entropy that have not been shown in the Letter.

	\subsection{Entanglement entropy for fixed commensurate density and $U_{\mathrm{LR}}=0.3U_0$}
	In this section we want to discuss the behavior of the entanglement entropy for fixed density and $U_{\mathrm{LR}}/U_0=0.3<0.5$ where the MI phase exists. We focus in this section on the continuous  transitions MI/SF and CDW/SS/SF when the density is fixed and commensurate by $\rho=1$ and $\rho=0.5$.
	
\begin{figure}[h!!!]
		\center
		{\includegraphics[width=1\linewidth]{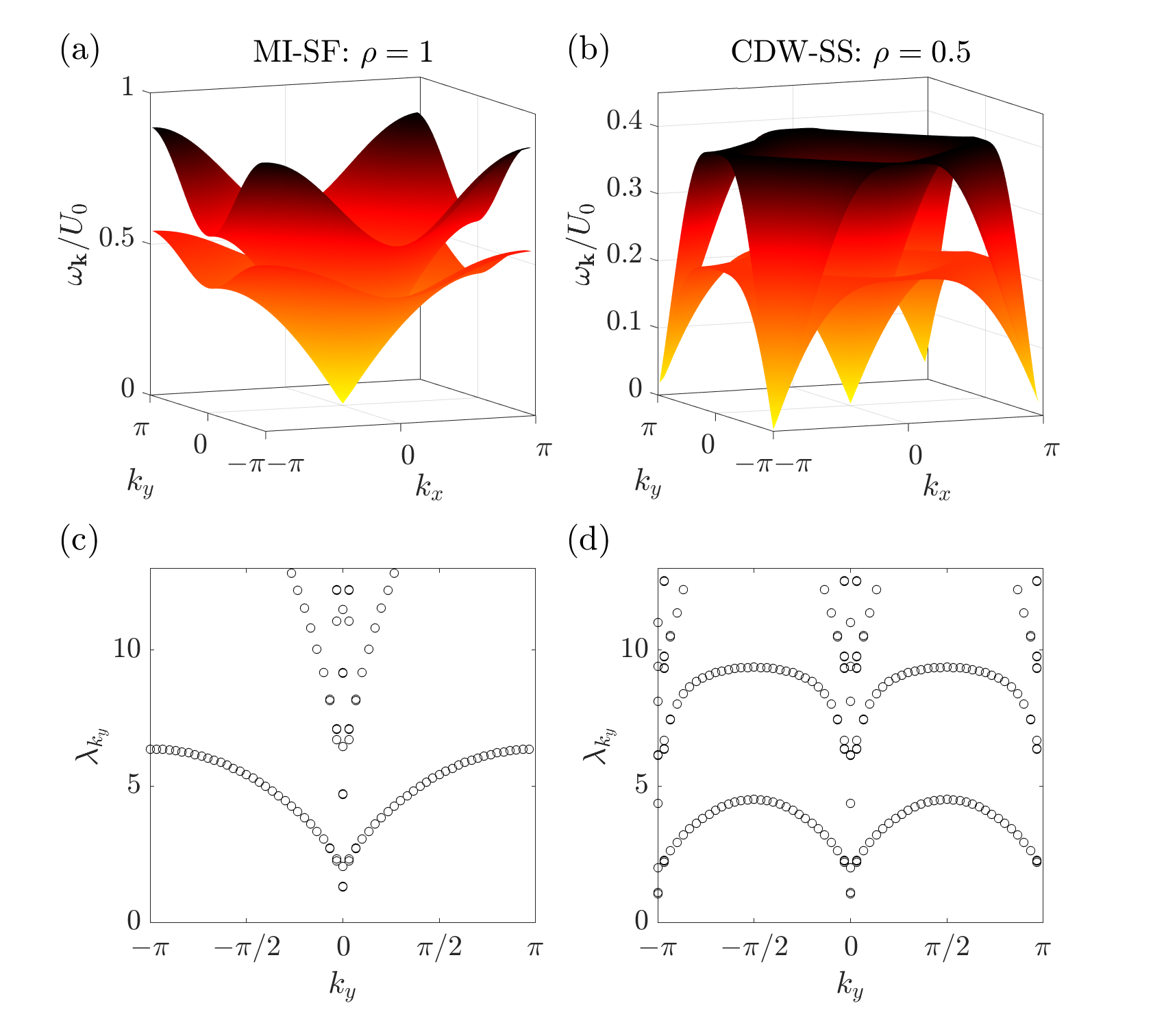}}
		\caption{The spectrum as function of the ${\bm k}$ wavevector at the transition from (a) MI/SF for $\rho=1$ ($t=0.17155U_0$, $\mu=0.4142U_0$)  and from (b) CDW/SF for $\rho=0.5$ ($t=0.23125U_0$, $\mu=-0.08632U_0$). The remaining parameters are $U_{\mathrm{LR}}=0.3U_0$, $L=60$. In (c) and (d) we show the entanglement spectrum as function of $k_y$ for the same parameter choices as in (a) and (b), respectively. For the calculation of the entanglement spectrum we have divided the system $L\times L$ into two sub-systems of size $L/2\times L$. } \label{FigS:2}
	\end{figure}

	As stated in the main text, we find that the mechanisms that leads to the singularity of the entanglement entropy at both transitions, MI/SF and CDW/SS, are very similar. The origin for the strong enhancement is the closing of the gap of the amplitude (Higgs-like) mode at the transition. 
	This is visible in Fig.~S\ref{FigS:2} where we show the energy spectrum at the (a) MI/SF and (b) CDW/SS transition for fixed density $\rho=1$ and $\rho=0.5$, respectively.  
	
	Both Figs.~S\ref{FigS:2}(a) and (b) show two gapless modes at the origin, corresponding to the Goldstone mode (lower branch) and the amplitude/Higgs mode (upper branch). For the CDW/SS transition we find the double gapless mode repeated at ${\bm k}=\boldsymbol{\pi}$ because of Brillouin-zone folding, as discussed in Sec.~\ref{kreprent}. In the entanglement spectrum [Figs.~S\ref{FigS:2}(c) and (d)] we find similar features that are here shown for the same parameter choices as in Figs.~S\ref{FigS:2}(a) and (b). There, we see that the lowest entanglement eigenmodes belong not only to the lower branch of the dispersion relation, but also to higher ones, reflecting the amplitude-mode softening in the excitation spectrum, and leading to a cusp singularity in the entanglement entropy  \cite{Frerot:2016}.
	
\begin{figure}[h!!!]
		\center
		{\includegraphics[width=1\linewidth]{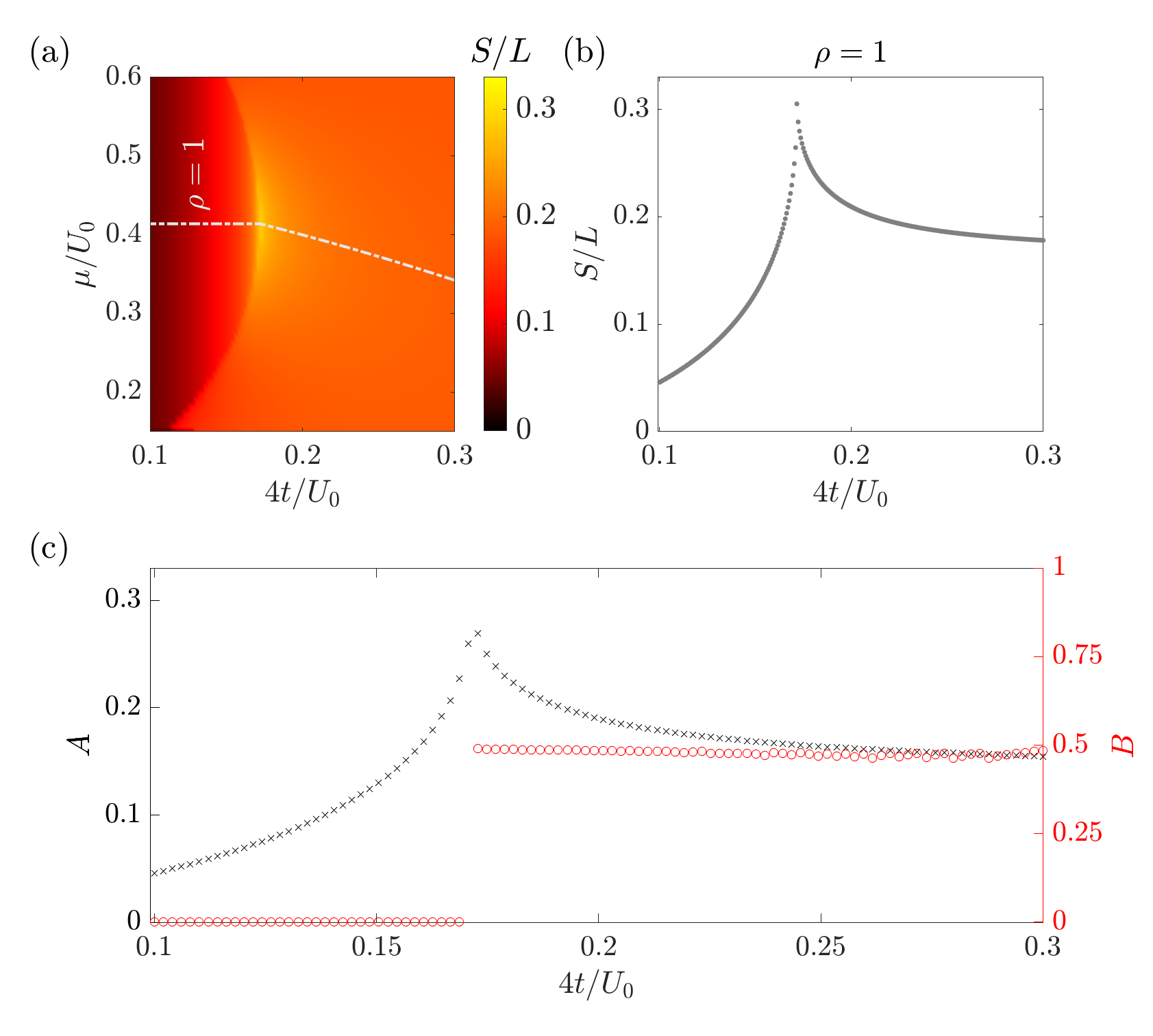}}
		\caption{(a) Entanglement entropy as function of $\mu$ and $t$ in units of $U_0$. The light gray dashed-dotted line show the parameters for which we find constant density $\rho=1$. (b) Entanglement entropy along a cut of fixed density for $\rho=1$ as function of $t$ in units of $U_0$. For (a) we have used $L=40$ and for (b) $L=60$ and $U_{\mathrm{LR}}=0.3U_0$. (c) The area law coefficient $A$ and the $\log$-correction $B$ as function of $t$ in units of $U_0$ across the MI/SF transition for fixed density $\rho=1$. The coefficients $A$ and $B$ are obtained by fitting the entanglement entropy with Eq.~\eqref{S:scal_ee} and $L=40,50,60,70,80,90,100,120,140$. } \label{FigS:3}
	\end{figure}

	In Fig.~S\ref{FigS:3}(a) we show the contour plot of the entanglement entropy as function of $\mu$ and $t$ in units of $U_0$. For a given density we find the corresponding $\mu$ value and draw the functions $\mu$ of $t$ visible as the  dashed-dotted line for density $\rho=1$. Along this line we see the enhanced singularity at the MI/SF transition visible in Fig.~S\ref{FigS:3}(b). The features are the same as dicussed in Ref.~\cite{Frerot:2016}.
	
	We calculate the entanglement entropy $S$ for various system sizes $L$ and fit $S(L)$ with
	\begin{align}
	S= AL+B\log{L}+C,\label{S:scal_ee}
	\end{align}
	where we extract the coefficients $A$, $B$, and $C$. The resulting coefficients $A$ and $B$ corresponding to the area law and the $\log$-correction are visible in Fig.~S\ref{FigS:3}(c). At the transition we see a spike in $A$ and a jump from $B\approx0$ (MI) to $B\approx N_G/2=0.5$ (SF), where $N_G$ is the number of Goldstone modes. This behavior is almost identical to the transition CDW/SS as we show in the following.
	
\begin{figure}
		\center
		{\includegraphics[width=1\linewidth]{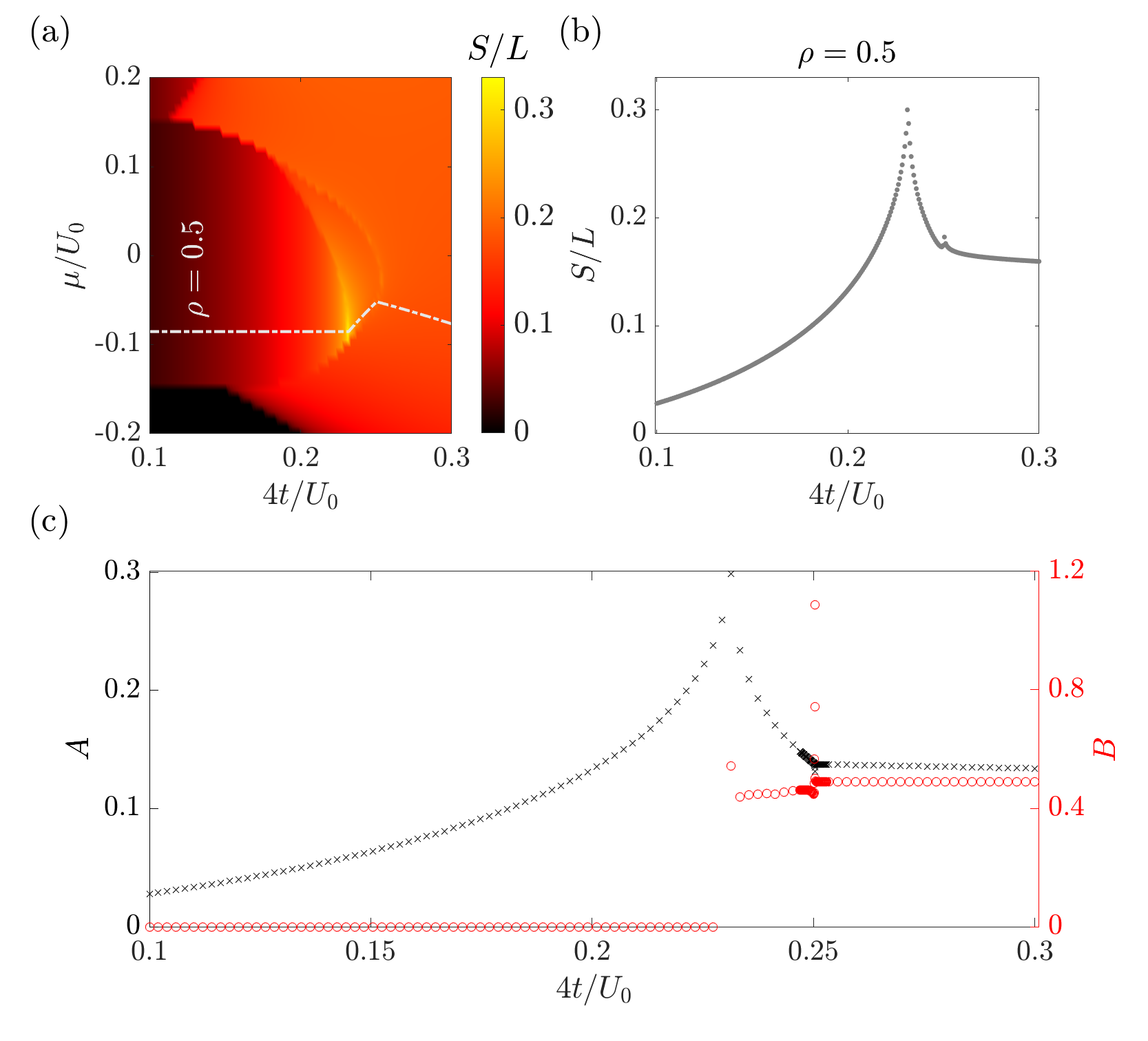}}
		\caption{(a) Entanglement entropy as function of $\mu$ and $t$ in units of $U_0$. The light gray dashed-dotted line show the parameters for which we find constant density $\rho=0.5$. (b) Entanglement entropy along a cut of fixed density for $\rho=0.5$ as function of $t$ in units of $U_0$. For (a) we have used $L=40$ and for (b) $L=60$ and $U_{\mathrm{LR}}=0.3U_0$. (c) The area law coefficient $A$ and the $\log$-correction $B$ as function of $t$ in units of $U_0$ for fixed density $\rho=0.5$. The coefficients $A$ and $B$ are obtained by fitting the entanglement entropy with Eq.~\eqref{S:scal_ee} and $L=40,50,60,70,80,90,100,120,140$. \label{FigS:7}} 
	\end{figure}

	In Fig.~S\ref{FigS:7}(a), we show now the trajectory in the $(t/U_0,\mu/U_0)$ plane that corresponds to a density $\rho=0.5$. This line crosses the continuous CDW/SS and SS/SF transitions. The entanglement entropy, visible in Fig.~S\ref{FigS:7}(b), shows a cusp at the CDW/SS that originates from the same spectral features as the MI/SF transition that has been obtained for $\rho=1$ [see Figs.~S\ref{FigS:2}(a), (b) and S\ref{FigS:3}(b)]. In addition, we also find the narrow spike of the entanglement entropy at the SS/SF transition, which is, however, far less pronounced than the one at the CDW/SS transition.
	
	In Fig.~S\ref{FigS:7}(c), we show the obtained values of $A$ and $B$ corresponding to the area law and the $\log$-correction. At the CDW/SS transition we find a sudden jump of $B$ to a value $B\sim0.5$ while the area law $A$ shows a spike. In fact the deviations from $B\sim0.5$ are visibly larger than they where in Fig~S\ref{FigS:3}(c). We can imagine that this is due to the very narrow parameter region where the SS phase is stable such that the entanglement entropy is still influenced by the nearby transition points. At the SS/SF transition, the $A$ coefficient shows a smooth behavior, while we find a narrow and high spike of the $\log$-correction $B$.

	\subsection{Entanglement entropy for incommensurate densities and $U_{\mathrm{LR}}=0.6U_0$}
	In this section we show additional results for the transition SS/SF at incommensurate densities for $U_{\mathrm{LR}}=0.6U_0$. We also show the results for the coefficients $A$ and $B$ describing the area law and $\log$-correction for incommensurate densities and $\mu=-0.05$ across the SS/SF and the CDW/SS transitions.
	
	\subsubsection{Roton mode in the excitation and entanglement spectrum}
	
\begin{figure*}[ht!]
		\center
		\includegraphics[width=0.8\linewidth]{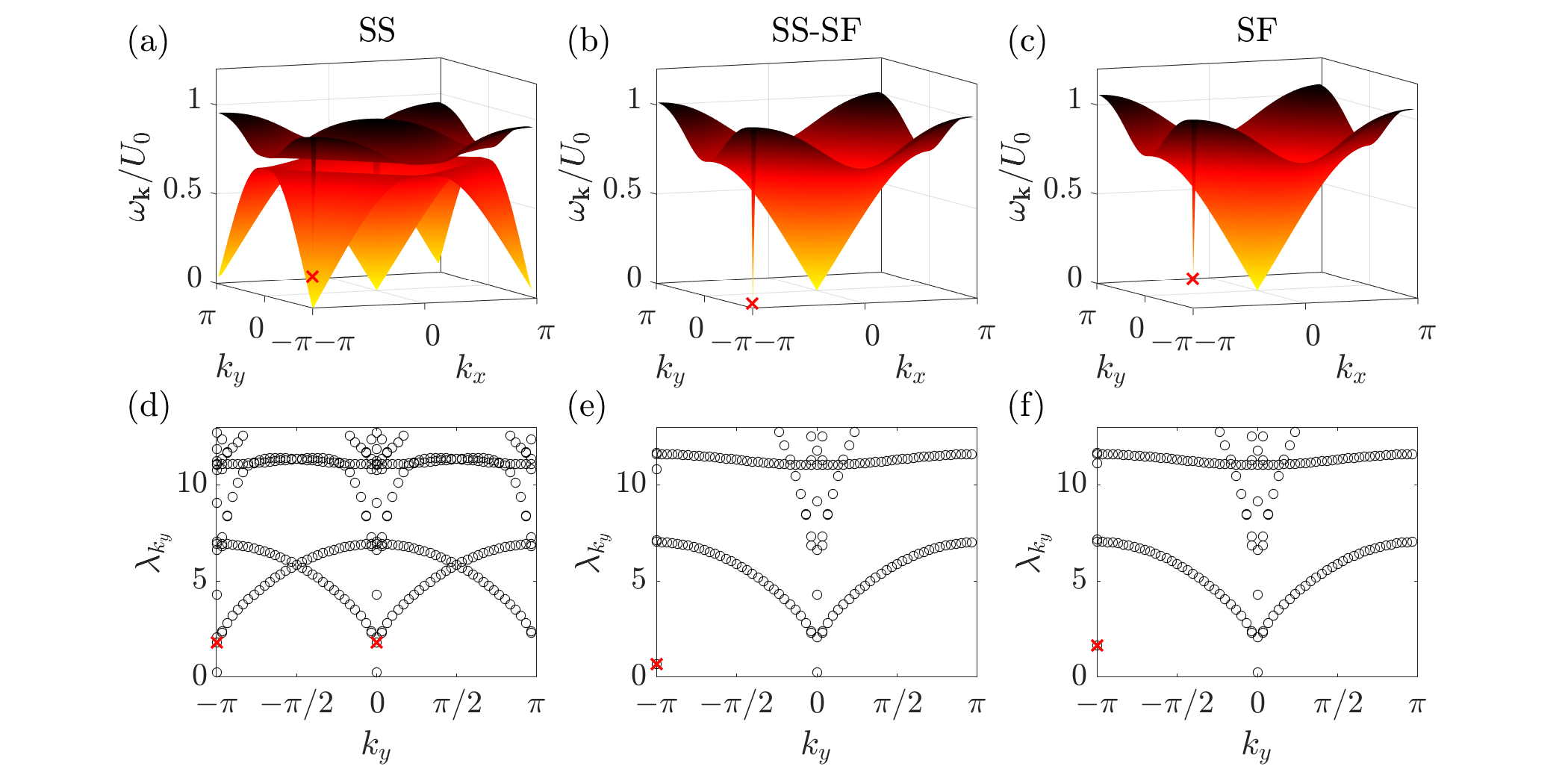}
		\caption {The excitation spectrum (a)-(c) and the entanglement spectrum (d)-(f) as function of the wavevector ${\bm k}=(k_x,k_y)$ and $k_y$, respectively. For all plots we have used a square lattice $L\times L$ with $L=60$. The entanglement spectrum is calculated by cutting the square lattice in two $L/2\times L$ sublattices. The red crosses are pointing at the roton mode. The spectra are calculated for $\mu=-0.1U_0$, $U_{\mathrm{LR}}=0.6U_0$, and for $4t=0.53U_0$ in the SS phase (a),(d), for $4t=0.542U_0$ at the transition from the SS to SF phases (b),(e), and for $4t=0.56U_0$ in the SF phase (c),(f). At the transition we see that the roton mode becomes gapless giving rise to the spike in the entanglement entropy.}\label{FigS:4}
	\end{figure*}

	First, we provide further details about the underlying nature of the energy and entanglement spectrum when crossing the SS/SF transition. In the main text we reported a narrow singular spike at the SS/SF transition that originates from the closing of the roton mode. In Fig.~S\ref{FigS:4} we show  the closing of the gap for this roton mode in the excitation spectrum (a)-(c) and also in the entanglement spectrum (d)-(f). In the SS phase [Fig.~S\ref{FigS:4}(a)] we show the dispersion relation across the complete Brillouin zone -- in order to clearly reveal the presence of the roton mode --  even though the actual Brillouin zone is reduced because of the appearance of the even-odd imbalance, as discussed in Sec.~\ref{kreprent}. 
	
	In the SS phase (Figs.~S\ref{FigS:4}(a) and (d)) we find the gapless Goldstone mode and a roton mode with a small but finite gap. The roton mode is highlighted by a red cross in Fig.~S\ref{FigS:4}(a). This feature is visible in the excitation and entanglement spectrum. 
	As already mentioned, because of the folding of the Brillouin zone onto a reduced one the roton mode is in fact properly sitting at ${\bm k}=0$ (excitation spectrum) and $k_y=0$  (entanglement spectrum), and it is repeated at the Brillouin-zone edge for the purpose of illustration.
	Approaching the transition SS/SF we observe that this roton mode becomes almost gapless in the excitation and the entanglement spectrum [Figs.~S\ref{FigS:4}(b) and (e)]. At the transition point the even-odd symmetry is restored, and the spectrum ``unfolds" over the entire Brillouin zone. Consequently, the roton mode is only visible at ${\bm k}=\boldsymbol{\pi}$ in the excitation spectrum and at $k_y=\pi$ in the entanglement spectrum.  Beyond the transition point, in the superfluid phase, we find again a gapped roton mode [Figs.~S\ref{FigS:4}(c) and (d)]. We emphasize that while the roton mode becomes gapless at the SS/SF transition and has a finite non-vanishing gap in the SS and SF phases, there is in addition always a gapless Goldstone mode. Therefore, while the area law of entanglement scaling is dominated by the gapless Goldstone mode, the critical behavior of the entanglement entropy across the SS/SF transition comes mostly from the closing of the roton gap, and it leads to the characteristic singularity in the prefactor of the subdominant logarithmic correction.
	
	\subsubsection{Incommensurate transition driven by the chemical potential}

\begin{figure}[h!]
		\center
		{\includegraphics[width=0.8\linewidth]{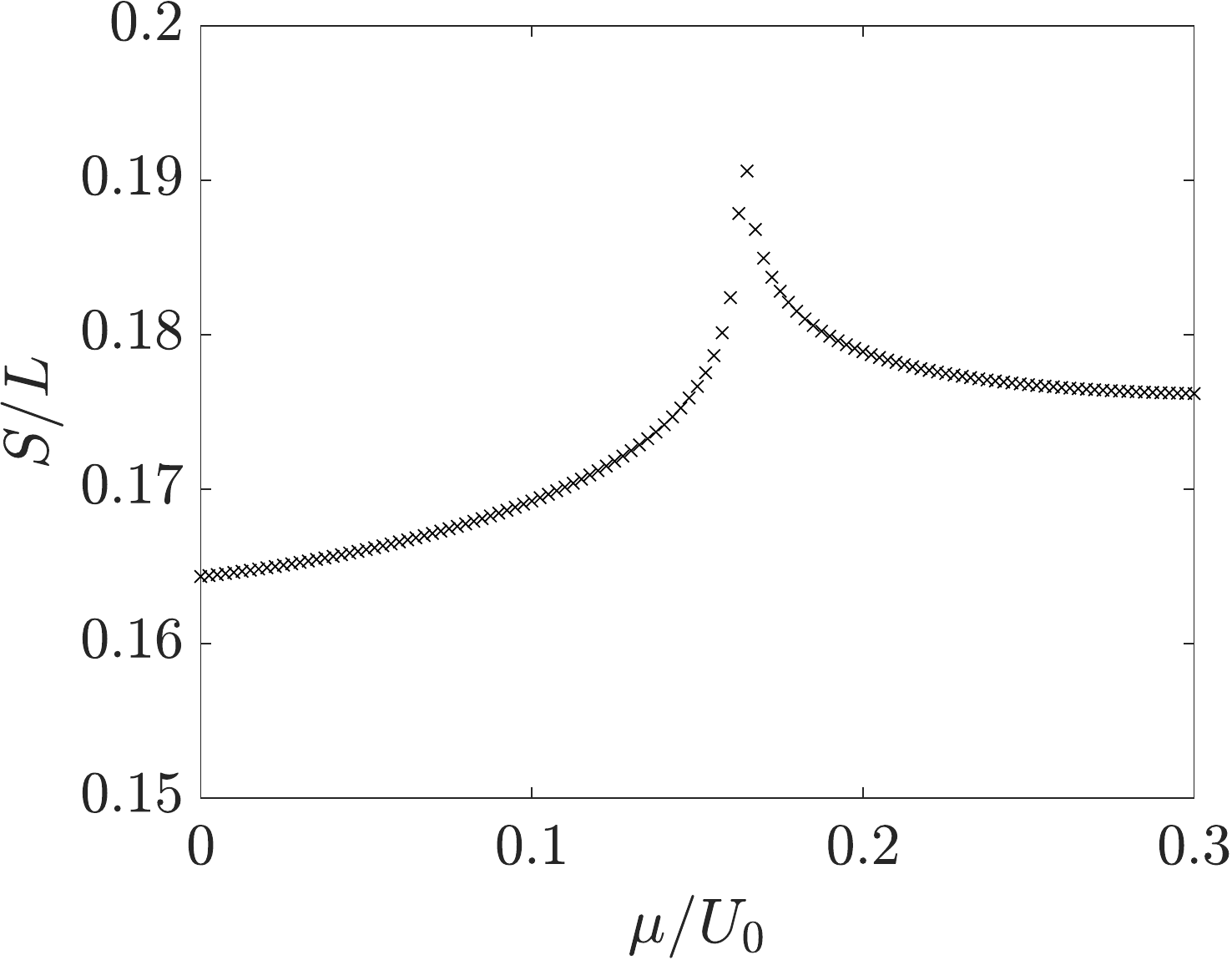}}
		\caption{Entanglement entropy as function of $\mu$ in units of $U_0$ across the SS/SF transition. The remaining parameters are $L=60$, $U_{\mathrm{LR}}=0.6U_0$, and $4t=0.5U_0$.} \label{FigS:5}
	\end{figure}

	To give further details on the incommensurate SS/SF transition, we consider the case in which the transition is crossed at fixed $t/U_0$ upon changing $\mu/U_0$. This, as shown in Fig.~S\ref{FigS:5}, leads to the same singular behavior of the entanglement entropy as the one seen before -- with the advantage that, along this cut across the phase diagram,  the SS/SF transition appears as isolated. The calculation shown in the figure corresponds to $U_{\mathrm{LR}}=0.6U_0$ and $4t=0.5U_0$. A similar behavior is found generically at the incommensurate SS/SF transition provided that it is continuous.

	\subsubsection{Incommensurate transition driven by the hopping}	
	
\begin{figure}[h!!!]
		\center
		{\includegraphics[width=1\linewidth]{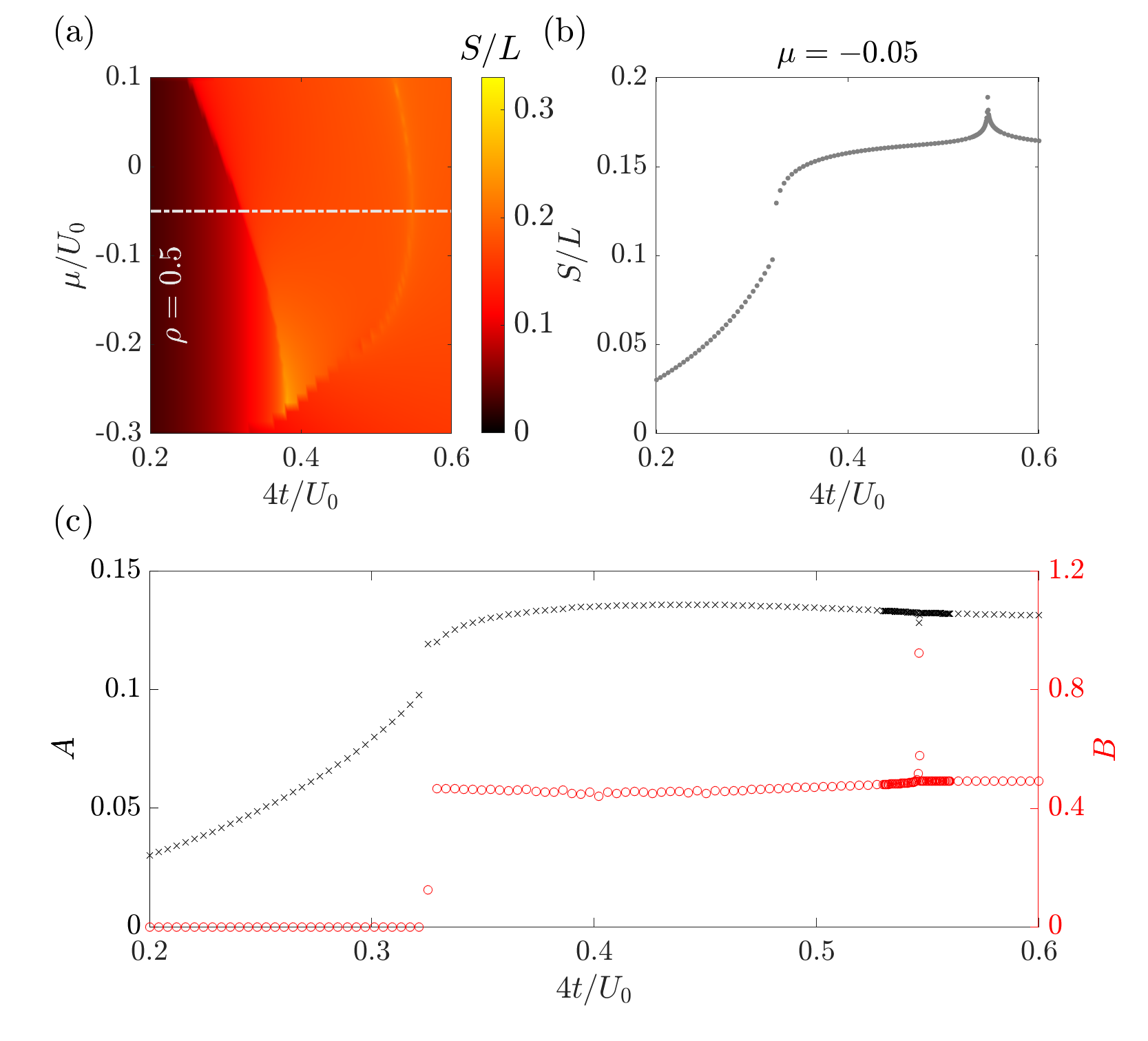}}
		\caption{(a) Entanglement entropy as function of $\mu$ and $t$ in units of $U_0$. The light gray dashed-dotted line corresponds to $\mu=-0.05U_0$. (b) Entanglement entropy along a cut for $\mu=-0.05U_0$ as function of $t$ in units of $U_0$. For (a) we have used $L=40$ and for (b) $L=60$ and $U_{\mathrm{LR}}=0.6U_0$. (c) The area law coefficient $A$ and the $\log$-correction $B$ as function of $t$ in units of $U_0$ for fixed $\mu=-0.05U_0$. The coefficients $A$ and $B$ are obtained by fitting the entanglement entropy with Eq.~\eqref{S:scal_ee} $L=40,50,60,70,80,90,100,120,140$. \label{FigS:8}} 
	\end{figure}

	To gain a complete picture, we also show data encompassing both the SS/SF transition and the CDW/SS transition driven by the $t/U_0$ ratio for fixed $\mu=-0.05U_0$. This data is supplementary to what is shown in the main text, where we only focus on the SS/SF transition. In Fig.~S\ref{FigS:8}(a) we show the transitions that we cross for fixed $\mu=-0.05U_0$, and Fig.~S\ref{FigS:8}(b) shows the entanglement entropy $S$ along this cut. 
	In the case at hand, the CDW/SS transition occurs at incommensurate density, and it falls in the parameter range in which it acquires a first-order nature. At this transition we find a very sharp transition in $S$ which appears to be a discontinuous jump.

	\subsubsection{Scaling analysis of the entanglement entropy across the transition}		
	Analyzing the behavior of the entanglement entropy across all transitions for different system sizes $L$, we can extract the coefficients $A$ and $B$ using Eq.~\eqref{S:scal_ee} corresponding to the area law and the $\log$-correction. The values obtained from this analysis are shown in Fig.~S\ref{FigS:8}(c). We have already discussed the narrow spike at the SS/SF transition that is shown in the Letter for exactly the same data set. 
	
	In the SS phase we find a nearly constant value of $B\approx0.5$, consistent with the presence of one gapless Goldstone mode, while this value drops to zero in the CDW phase, consistent with the disappearance of the Goldstone mode and the restoration of the $U(1)$ symmetry.
	
	\subsection{Modifications of the $\log$ correction for different regularization}
	
	In this section we show that the spike feature in the prefactor of the logarithmic correction at the SS/SF transition is robust against a modification of the regularizing field $h(L)$.
	
	As already mentioned in the main text, the appearance of strictly zero-energy modes within a quadratic theory on finite-size systems is unphysical -- as a finite-size gap is always to be expected -- and it therefore needs regularization via the application of external fields. To this scope we modify the local Hamiltonian given in Eq.~\eqref{Hloc} by
	\begin{align}
	\hat{H}'_{\bm r}=\hat{H}_{\bm r}-g(L)(\hat{b}_{\bm r}^{\dag}+\hat{b}_{\bm r})-h(L)Z_{\bm r}\hat{n}_{\bm r}.\label{Hloc2}
	\end{align}
	introducing fields $g$ and $h$ which couple to the SF order parameter and to the crystalline order parameter, in order to gap out the Goldstone and roton mode respectively. 
	The first field $g$ scales as $g(L)\sim L^{-\tilde{\kappa}}$ with $\tilde{\kappa}=4$ such that the Goldstone excitation gap scales as $\lambda_{\mathrm{G}}\sim L^{-2}$: this scaling choice  provides the correct logarithmic contribution from the Goldstone mode to the entanglement entropy with prefactor $B=0.5$ in the SF phase as well as in the SS phase~\cite{Frerot:2016,Frerot:2015,Song:2010}.
	
	As for the choice of the scaling dependence of $h(L)$, for the results shown in the main text we have chosen the form $h(L)\sim L^{-\kappa}$ with $\kappa=4$  -- in analogy to the scaling of the $g(L)$ field. As stated in the main text, a more educated choice would require the knowledge of the dynamical critical exponent at the SS/SF transition (to be imitated by the scaling of the roton gap); yet this exponent is currently not known for the transition in question.  
	
\begin{figure}[t]
		\center
		\includegraphics[width=1\linewidth]{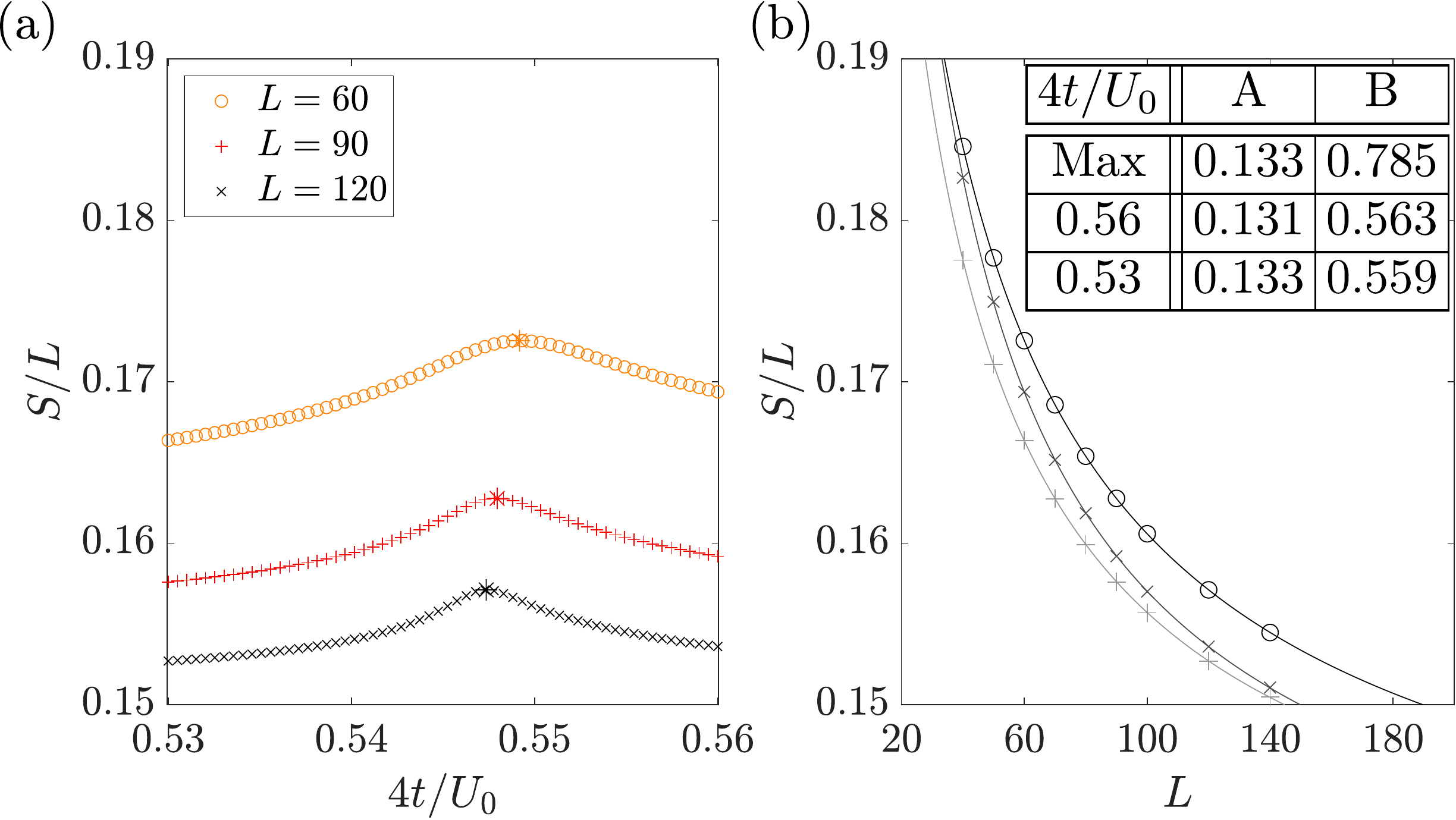}\
		\caption{(a) Entanglement entropy as function of the tunneling rate $t$ in units of $U_0$ across the transition from SS/SF, calculated by cutting a $L\times L$ lattice into two $L/2\times L$ lattices with different $L$ (see legend) for $U_\mathrm{LR}=0.6U_0$ and constant $\mu=-0.05U_0$. The ``$*$'' symbols show the maximum of $S$ in the given interval. (b) The maximum of $S$ (black ``o") and $S$ for $4t/U_0=0.56$ (dark gray ``x"), $4t/U_0=0.53$ (gray ``+") for different $L$.   The coefficients $A$ and $B$ are obtained by fitting Eq.~\eqref{S:scal_ee} to $S$ and are given in the table. }\label{Fig:S6}
	\end{figure}

	In spite of this apparent ambiguity, we can show that a modification of the $\kappa$ exponent does not alter the main conclusion of our work, namely the presence of a singularity in the logarithmic correction to the scaling of the entanglement entropy at the SS/SF transition. To show this explicitly, we calculate the coefficients $A$ and $B$ for the same SS/SF transition discussed in the main text, but this time using $\kappa=2$. The results are shown in Fig.~S\ref{Fig:S6}.
	For a reduced value of $\kappa$ we expect also a smaller exponent for the closing of the roton mode $\lambda_{\mathrm{rot}}$. Consequently, we find a more ``rounded'' entanglement entropy $S$ as shown in Fig.~S\ref{Fig:S6}(a). Nevertheless, fitting the entanglement entropy at its maximum results in a much larger $B$ coefficient than away from the critical point (see table in Fig.~S\ref{Fig:S6}(b)). This shows that while the coefficient at the maximum is here smaller than for $\kappa=4$ ($B\approx 0.8$ instead of  $B\approx0.9$) it is still significantly pronounced with respect to the value $B\approx0.5$ that is expected and found in the SS and SF phases away from the critical point.

	\section{Slave-boson theory for the quantum critical scaling of entanglement in the Lipkin-Meshkov-Glick model}
	\begin{figure*}[ht!]
		\begin{center}
			\includegraphics[width=0.85\textwidth]{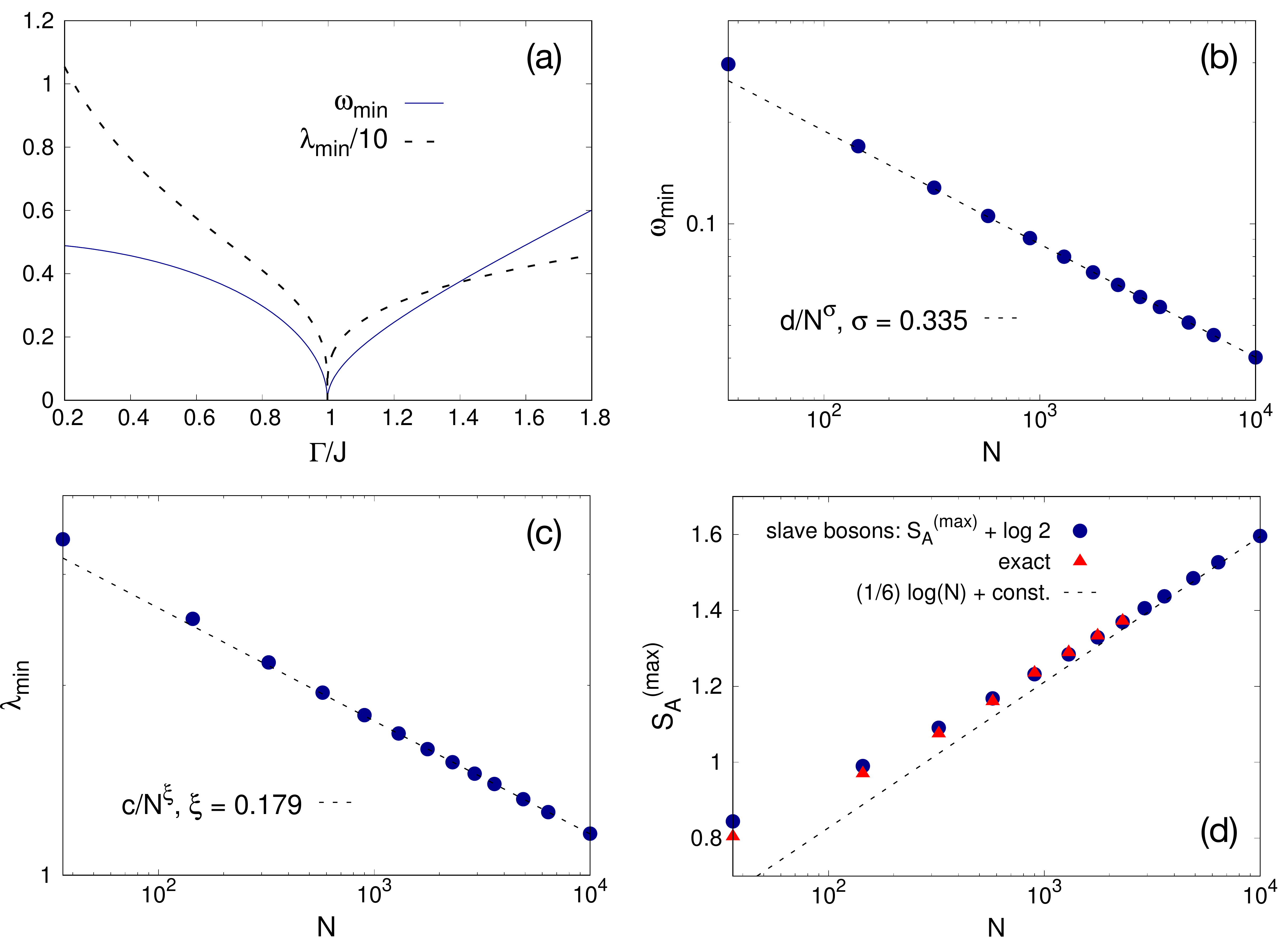}
			\caption{(a) Slave-boson prediction for the minimal excitation frequency ($\omega_{\rm min}$) and minimal entanglement frequency ($\lambda_{\rm min}$) across the quantum phase transition of the LMG model (system size $N=400$); (b) scaling of the minimal excitation frequency $\omega_{\rm min}$ at the finite-size/finite-$h$ critical field $\Gamma_c(N;h(N))$, at which $\omega_{\rm min}$ is minimal as a function of $\Gamma$ -- the dashed line is a power-law fit; (c) scaling of the minimal entanglement frequency at the finite-size/finite-$h$ critical field $\Gamma_c(N;h(N))$ -- same significance for the dashed line as in the previous panel; (d) scaling of the half-system entanglement entropy at $\Gamma_c(N;h(N))$ from slave-boson theory, compared with the exact results for the LMG model (taken at a different size-dependent field $\Gamma'_c(N)$, corresponding to the maximum entropy for the exact solution as a function of $\Gamma$); and with the expected exact behavior at large size $S = (1/6) \log N + const.$.} \label{FigLMG}
		\end{center}
	\end{figure*}
	
	In this section we discuss the slave-boson treatment of the $\mathbb{Z}_2$ quantum phase transition exhibited by the Lipkin-Meshkov Glick model \cite{Lipkin:1965,Meshkov:1965,Glick:1965}, which corresponds to the Ising model in a transverse field with infinite-range ferromagnetic couplings.  We show that: 1) the slave-boson theory with appropriate regularization of the zero mode captures exactly the scaling of the entanglement entropy at the quantum phase transition (logarithmic scaling behavior with the correct prefactor); and 2) that the scaling of the entanglement entropy at the transition is dominated by the singular vanishing of an isolated mode in the entanglement spectrum, reflecting a similar mode in the excitation spectrum. This behavior is fully analogous to that exhibited by the extended Bose-Hubbard model with infinite-range interactions at the $\mathbb{Z}_2$ supersolid/superfluid transition, as detailed in the main text.
	
	The Hamiltonian for the long-range (LR) Ising model in a transverse ($\Gamma$) and longitudinal ($h$) field reads
	\begin{equation}
	\hat{H}_{\rm LR-Ising}  = -\frac{1}{N} \sum_{ij} J_{ij} \hat s_i^z \hat s_j^z - \Gamma \sum_i \hat s_i^x - h  \sum_i \hat s_i^z
	\end{equation}
	where $\hat s_i^\alpha$ ($\alpha = x, y, z$) are $s=1/2$ spin operators, and $J_{ij} = J/|\bm r_i-\bm r_j|^\alpha$ (with $J>0$) describes a ferromagnetic, power-law decaying interaction. The $\alpha=0$ limit corresponds to the LMG model. The mean-field solution for the ground state of the above model reads ${|\Psi_{0,\rm MF}\rangle = [\cos(\theta/2) |\uparrow\rangle + \sin(\theta/2) |\downarrow\rangle]^{\otimes^N}}$, and the energy minimization condition for the $\theta$ angle reads
	\begin{equation}
	\chi_0  \cos\theta\sin\theta - \Gamma\cos\theta + H \sin\theta = 0~.
	\end{equation}
	Here we have introduced the function $\chi_{\bm k} = \frac{1}{N^2} \sum_{ij} e^{i{\bm k}\cdot(\bm r_i - \bm r_j)} J_{ij}$, corresponding to the Fourier transform of the interactions. In the case of the LMG model $\chi_{\bm k} = \chi_0 \delta_{\bm k,0}$ with $\chi_0 = J$. For $h=0$ the system has a $\mathbb{Z}_2$ quantum phase transition at $\Gamma_c = J$, separating a ferromagnetic ground state ($\Gamma/J<1$) from a paramagnetic one ($\Gamma/J>1$) -- given the collective nature of the interactions, the mean-field prediction for the position of the transition is exact in the thermodynamic limit. 
	
	In order to investigate harmonic quantum fluctuations around the mean-field solution, one should first rotate the spins by an angle $\theta$ around the $y$ axis, so as to align the quantization axis with the direction of the mean-field ground state. This defines then the new operators $\hat s^{z'}_j$ and $\hat s^{x'}_j$, such that 
	\begin{eqnarray}
	\hat s^{z}_j & = & \cos\theta~ \hat s^{z'}_j - \sin\theta~ \hat s^{x'}_j  \nonumber \\
	\hat s^{x}_j & = & \sin\theta~ \hat s^{z'}_j + \cos\theta ~\hat s^{x'}_j~.
	\end{eqnarray}
	These spin operators can then mapped onto Schwinger (slave) bosons
	\begin{eqnarray}
	\hat s^{z'}_j & = &  \frac{1}{2} \left (  \hat a^\dagger_j \hat a_j - \hat b^\dagger_j \hat b_j \right)   \nonumber \\
	\hat s^{x'}_j &=& \frac{1}{2} \left ( \hat a^\dagger_j \hat b_j + \hat b^\dagger_j \hat a_j \right )  
	\end{eqnarray}
	with the constraint $\hat a^\dagger_j \hat a_j +  \hat b^\dagger_j \hat b_j =1$, where $\hat a_j = \hat \gamma_{j,\uparrow}$ and $\hat b_j = \hat \gamma_{j,\downarrow}$ to adopt the notation used in the main text. The harmonic treatment of quadratic fluctuations around the mean-field state amounts then to take $\hat a_j, \hat a^\dagger_j \approx 1$ and $\langle \hat b^\dagger_j \hat b_j \rangle \ll 1$, allowing for a quadratic expansion of the Hamiltonian in the $\hat b_j, \hat b^\dagger_j$ operators; under these assumptions one recovers the same quadratic bosonic Hamiltonian as that of linear spin-wave theory (based on the linearized Holstein-Primakoff spin-boson mapping). 
	
	Employing a Fourier transformation the Hamiltonian $\hat{H}$ can then be written in the form $\hat{H} \approx \langle \hat H \rangle_{\rm MF} + \hat H^{(2)}$ where 
	\begin{eqnarray}
	\hat H^{(2)} & = &  \frac{1}{2} \sum_{\bm k} {\cal A}_{\bm k} \left ( \hat b^\dagger_{\bm k} \hat b_{\bm k} + \hat b^\dagger_{-\bm k} \hat b_{-\bm k} \right ) \nonumber \\
	&+&  \frac{1}{2} \sum_{\bm k} {\cal B}_{\bm k} \left ( \hat b^\dagger_{\bm k} \hat b^\dagger_{-\bm k} + {\rm h.c.} \right ) 
	\end{eqnarray}
	and 
	\begin{eqnarray}
	{\cal A}_{\bm k} &=& \chi_0 \cos^2 \theta + \Gamma \sin\theta - \frac{1}{2} \chi_{\bm k} \sin^2\theta + h \cos\theta \nonumber \\
	{\cal B}_{\bm k} & = &  - \frac{1}{2} \chi_{\bm k} \sin^2\theta~.
	\end{eqnarray} 
	The quadratic Hamiltonian can be Bogolyubov diagonalized with the linear transformation $\hat b_{\bm k} = u_{\bm k} \hat c_{\bm k} - v_{\bm k} \hat c^\dagger_{-\bm k}$, which leads to the form 
	\begin{equation}
	\hat{H}^{(2)} = \sum_{\bm k} \omega_{\bm k} \hat c^\dagger_{\bm k} \hat c_{\bm k} + {\rm const.}
	\end{equation} 
	with $\omega_{\bm k} = \sqrt{ {\cal A}_{\bm k}^2 - {\cal B}_{\bm k}^2 }$, provided that $u_{\bm k} = \sqrt{({\cal A}_{\bm k}/\omega_{\bm k} + 1)/2}$ and $v_{\bm k} = \sqrt{({\cal A}_{\bm k}/\omega_{\bm k} - 1)/2}$. 
	Due to the collective nature of the interactions, the excitation spectrum $\omega_{\bm k}$ is dispersionless except at ${\bm k}=0$, where it exhibits a soft mode. Such a mode is gapped for all values of $\Gamma$ except for $\Gamma_c(N) = \Gamma_c + \Delta \Gamma(N)$ (where $\Delta\Gamma(N)$ is a finite-size correction vanishing with $N$), at which the frequency vanishes strictly for all system sizes $N$ when $h=0$ -- see Fig.~S\ref{FigLMG}(a). The application of a small, size-dependent field $h=h(N)$ allows one to open an equally size-dependent gap, mimicking the one appearing in the exact solution of the problem; the minimal-gap field is also shifted to $\Gamma_c(N;h) = \Gamma_c + \Delta \Gamma(N;h)$. The finite-size gap in the exact solution of the LMG model is known to scale as $\Delta \sim N^{-1/3}$ at criticality \cite{Vidal:2007}. As shown in Fig.~S\ref{FigLMG}(b), the application of a field $h = J/N$ leads precisely to this scaling of the gap for the harmonic excitations. 
	
	The harmonic ground state corresponds to the vacuum of the $\hat c$ bosons. The covariance matrix of the $\hat b$ bosons for the half-system $\mathbb{A}$ can then be readily calculated for such a state, and from its diagonalization we can obtain the entanglement spectrum and the entanglement entropy in the same way as for the extended BH model. Similarly to the excitation spectrum, the (one-body) entanglement spectrum  exhibits a unique soft mode at frequency $\lambda_{\rm min}$ which is always gapped except at the (finite-size-corrected) critical point $\Gamma_c(N)$, where it becomes strictly gapless if $h=0$ (see Fig.~S\ref{FigLMG}(a)); this is unphysical, as it leads to a strict divergence of the sub-system entanglement entropy  as $S \sim - \log \lambda_{\rm min}$ for any subsystem size. 
	
	On the other hand, the regularization of the zero mode in the excitation spectrum leads to a similar regularization of  $\lambda_{\rm min}$, which at criticality acquires a power-law scaling $\lambda_{\rm min} \sim N^{-\xi}$, as shown in Fig.~S\ref{FigLMG}(c), where $\xi \approx 0.179$. As a consequence one would conclude that, at criticality,  ${S = \xi \log N + ...}$, where $(...)$ stands for sub-leading terms. This logarithmic scaling is already in agreement with the exact solution for the scaling of the half-system entanglement entropy, which is predicted to grow as ${S = (1/6) \log N + ...}$~ \cite{Vidal:2007}. Yet the $\xi$ factor coming from the single soft mode differs from the $1/6$ factor expected from the exact solution.  Nonetheless, for large $N$ values the gapped entanglement modes around the soft mode are found to contribute not only to the subleading scaling terms, but also to the dominant logarithmic scaling. Extracting correctly the entanglement entropy from the entire entanglement spectrum, we find that the critical behavior of $S$ mimics in fact very closely that of the exact solution (see Fig.~S\ref{FigLMG}(d)), provided that the entropy obtained with the harmonic (slave-boson) theory is shifted by a factor $\log(2)$. The latter shift accounts for the fact that the harmonic theory explicitly breaks the $\mathbb{Z}_2$ symmetry, and therefore it misses half of the Hilbert space accessible to the system.  
	
	From the above results we can therefore conclude that the careful regularization of the zero mode in the harmonic theory allows one to recover the \emph{exact} scaling of the half-system entanglement entropy at the $\mathbb{Z}_2$ quantum critical point of the LMG model. This result leads us to believe that our slave-boson treatment of the  $\mathbb{Z}_2$ supersolid/superfluid transition in the presence of infinite-range interactions is also accurate; and that one can also reconstruct the exact value of the prefactor of the logarithmic term related to the critical softening of the roton mode, provided that one applies a regularizing $h$ field that produces the right scaling of the finite-size gap in the excitation spectrum at criticality.  
	
	\end{document}